\documentclass[american,aps,prl,reprint,superscriptaddress]{revtex4-1}
\usepackage[T1]{fontenc}
\usepackage[latin9]{inputenc}
\usepackage{color}
\usepackage{babel}
\usepackage{amsmath}
\usepackage{amsthm}
\usepackage{amssymb}
\usepackage{graphicx}
\usepackage[unicode=true,pdfusetitle,
 bookmarks=true,bookmarksnumbered=false,bookmarksopen=false,
 breaklinks=false,pdfborder={0 0 0},backref=false,colorlinks=true]
 {hyperref}

\makeatletter

\providecommand{\tabularnewline}{\\}

\theoremstyle{plain}
\newtheorem{thm}{\protect\theoremname}
  \theoremstyle{plain}
  \newtheorem{lem}[thm]{\protect\lemmaname}

\usepackage{babel}
\usepackage{pxfonts}
\usepackage{braket}
\usepackage{colortbl}
\definecolor{myurlcolor}{rgb}{0,0,0.7}

\providecommand{\theoremname}{Theorem}

\makeatother

  \providecommand{\lemmaname}{Lemma}
\providecommand{\theoremname}{Theorem}

\begin{document}

\title{Activation of entanglement from quantum coherence and superposition}

\author{Lu-Feng Qiao}

\thanks{These authors contributed equally to this work.}
\affiliation{State Key Laboratory of Advanced Optical Communication Systems and
Networks, Institute of Natural Sciences $\&$ Department of Physics
and Astronomy, Shanghai Jiao Tong University, Shanghai 200240, China}

\affiliation{Synergetic Innovation Center of Quantum Information and Quantum Physics,
University of Science and Technology of China, Hefei, Anhui 230026,
China}

\author{Jun Gao}

\thanks{These authors contributed equally to this work.}
\affiliation{State Key Laboratory of Advanced Optical Communication Systems and
Networks, Institute of Natural Sciences $\&$ Department of Physics
and Astronomy, Shanghai Jiao Tong University, Shanghai 200240, China}

\affiliation{Synergetic Innovation Center of Quantum Information and Quantum Physics,
University of Science and Technology of China, Hefei, Anhui 230026,
China}

\author{Alexander Streltsov}

\thanks{These authors contributed equally to this work.}
\affiliation{Faculty of Applied Physics and Mathematics, Gda\'{n}sk University
of Technology, 80-233 Gda\'{n}sk, Poland}

\affiliation{National Quantum Information Centre in Gda\'{n}sk, 81-824 Sopot,
Poland}

\author{Swapan Rana}

\affiliation{ICFO -- Institut de Ciencies Fotoniques, The Barcelona Institute
of Science and Technology, ES-08860 Castelldefels, Spain}

\author{Ruo-Jing~Ren}

\affiliation{State Key Laboratory of Advanced Optical Communication Systems and
Networks, Institute of Natural Sciences $\&$ Department of Physics
and Astronomy, Shanghai Jiao Tong University, Shanghai 200240, China}

\affiliation{Synergetic Innovation Center of Quantum Information and Quantum Physics,
University of Science and Technology of China, Hefei, Anhui 230026,
China}

\author{Zhi-Qiang Jiao}

\affiliation{State Key Laboratory of Advanced Optical Communication Systems and
Networks, Institute of Natural Sciences $\&$ Department of Physics
and Astronomy, Shanghai Jiao Tong University, Shanghai 200240, China}

\affiliation{Synergetic Innovation Center of Quantum Information and Quantum Physics,
University of Science and Technology of China, Hefei, Anhui 230026,
China}

\author{Cheng-Qiu Hu}

\affiliation{State Key Laboratory of Advanced Optical Communication Systems and
Networks, Institute of Natural Sciences $\&$ Department of Physics
and Astronomy, Shanghai Jiao Tong University, Shanghai 200240, China}

\affiliation{Synergetic Innovation Center of Quantum Information and Quantum Physics,
University of Science and Technology of China, Hefei, Anhui 230026,
China}

\author{Xiao-Yun Xu}

\affiliation{State Key Laboratory of Advanced Optical Communication Systems and
Networks, Institute of Natural Sciences $\&$ Department of Physics
and Astronomy, Shanghai Jiao Tong University, Shanghai 200240, China}

\affiliation{Synergetic Innovation Center of Quantum Information and Quantum Physics,
University of Science and Technology of China, Hefei, Anhui 230026,
China}

\author{Ci-Yu Wang}

\affiliation{State Key Laboratory of Advanced Optical Communication Systems and
Networks, Institute of Natural Sciences $\&$ Department of Physics
and Astronomy, Shanghai Jiao Tong University, Shanghai 200240, China}

\affiliation{Synergetic Innovation Center of Quantum Information and Quantum Physics,
University of Science and Technology of China, Hefei, Anhui 230026,
China}

\author{Hao Tang}

\affiliation{State Key Laboratory of Advanced Optical Communication Systems and
Networks, Institute of Natural Sciences $\&$ Department of Physics
and Astronomy, Shanghai Jiao Tong University, Shanghai 200240, China}

\affiliation{Synergetic Innovation Center of Quantum Information and Quantum Physics,
University of Science and Technology of China, Hefei, Anhui 230026,
China}

\author{Ai-Lin Yang}

\affiliation{State Key Laboratory of Advanced Optical Communication Systems and
Networks, Institute of Natural Sciences $\&$ Department of Physics
and Astronomy, Shanghai Jiao Tong University, Shanghai 200240, China}

\affiliation{Synergetic Innovation Center of Quantum Information and Quantum Physics,
University of Science and Technology of China, Hefei, Anhui 230026,
China}

\author{Zhi-Hao Ma}

\affiliation{Department of Mathematics, Shanghai Jiaotong University, Shanghai
200240, China}

\author{Maciej Lewenstein}

\affiliation{ICFO -- Institut de Ciencies Fotoniques, The Barcelona Institute
of Science and Technology, ES-08860 Castelldefels, Spain}

\affiliation{ICREA, Pg.~Lluis Companys 23, ES-08010 Barcelona, Spain}

\author{Xian-Min Jin}

\thanks{xianmin.jin@sjtu.edu.cn}
\affiliation{State Key Laboratory of Advanced Optical Communication Systems and
Networks, Institute of Natural Sciences $\&$ Department of Physics
and Astronomy, Shanghai Jiao Tong University, Shanghai 200240, China}

\affiliation{Synergetic Innovation Center of Quantum Information and Quantum Physics,
University of Science and Technology of China, Hefei, Anhui 230026,
China}


\maketitle
\textbf{Quantum entanglement and coherence are two fundamental features
of nature, arising from the superposition principle of quantum mechanics \cite{Schrodinger1935a}. While considered
as puzzling phenomena in the early days of quantum theory \cite{EinsteinPhysRev.47.777},
it is only very recently that entanglement and coherence have been
recognized as resources for the emerging quantum technologies, including
quantum metrology, quantum communication, and quantum computing \cite{Horodecki2009,Streltsov2016}.
In this work we study the limitations
for the interconversion between coherence and
entanglement. We prove a fundamental
no-go theorem, stating that a general resource theory of superposition
does not allow for entanglement activation. By
constructing a CNOT gate as a free operation, we experimentally
show that such activation is possible within the more constrained
framework of quantum coherence. Our results
provide new insights into the interplay between coherence and entanglement,
representing a substantial step forward for solving longstanding open questions in quantum
information science.}

Quantum resource theories provide a fundamental framework for studying
general notions of nonclassicality, including quantum entanglement
\cite{Vedral1997,Horodecki2009} and coherence \cite{Baumgratz2014,Streltsov2016}.
Any such resource theory is based on the notion of free states and
free operations. Free operations are physical transformations which
do not consume any resources. They strongly depend on the problem
under study, and are usually motivated by physical or technological
constraints. In entanglement theory, these constraints are naturally
given by the \emph{distance lab paradigm}: two spatially separated
parties can perform quantum measurements in their local labs, but
can only exchange classical information between each other.

Free states of a resource theory are quantum states which can be produced without consuming any resources.
In entanglement theory, these free states are called separable \cite{WernerPhysRevA.40.4277}.
Various quantum protocols require the presence
of entanglement. This includes quantum teleportation \cite{BennettPhysRevLett.70.1895,Jin2010}, quantum cryptography \cite{EkertPhysRevLett.67.661},
and quantum state merging \cite{Horodecki2005}. As has been demonstrated
very recently, it is indeed possible to establish and maintain high
degree of entanglement via large distances \cite{Yin2017}.

The resource theory of quantum coherence studies technological limitations
for establishing quantum superpositions \cite{Baumgratz2014,Streltsov2016}.
This theory requires the existence of a distinguished basis, which
can be interpreted as \emph{classical}, and is usually present due
to the unavoidable decoherence \cite{Zurek2003}. Quantum states belonging
to this basis are then called incoherent, and considered as the free
states of coherence theory. Superpositions of these free states are
said to possess coherence. Incoherent operations
are free operations of coherence theory: they correspond to quantum
measurements which do not create coherence for individual measurement
outcomes \cite{Baumgratz2014}. Recent results show that coherence
plays a crucial role for quantum metrology \cite{Giovannetti2011NaPho,MarvianPhysRevA.94.052324},
and that coherence might be more suitable than entanglement to capture
the performance of quantum algorithms \cite{HilleryPhysRevA.93.012111,Matera2016}.
Recent investigations also show that coherence and entanglement play
an important role in biological systems \cite{Huelga2013}.

Due to the aforementioned significance of coherence and entanglement
for quantum technologies, it is crucial to understand how these fundamental
resources can be converted into each other. In this work we address
this question, and confirm our theoretical results
by an experiment with photons. We present a fundamental no-go theorem,
showing that a general resource theory of superposition does not allow
for entanglement activation, while this is possible within the more
constrained theory of coherence. This result shares
the same spirit with the celebrated no-cloning theorem \cite{Wootters1982}:
a general quantum state cannot be copied, while cloning is in fact
possible for a restricted set of mutually orthogonal states. We experimentally
demonstrate entanglement activation from coherence by preparing photon
states with different degrees of coherence and activating them into
entanglement by applying an optical CNOT gate. Our
results lead to a fundamental insight about entanglement quantifiers,
proving that trace norm entanglement violates strong monotonicity.
This shows how recent results on the resource theory of quantum coherence
can be used for solving important open questions in quantum information
science.

\section*{No-go Theorem of Entanglement Activation}

\begin{figure}[h]
\includegraphics[width=0.95\columnwidth]{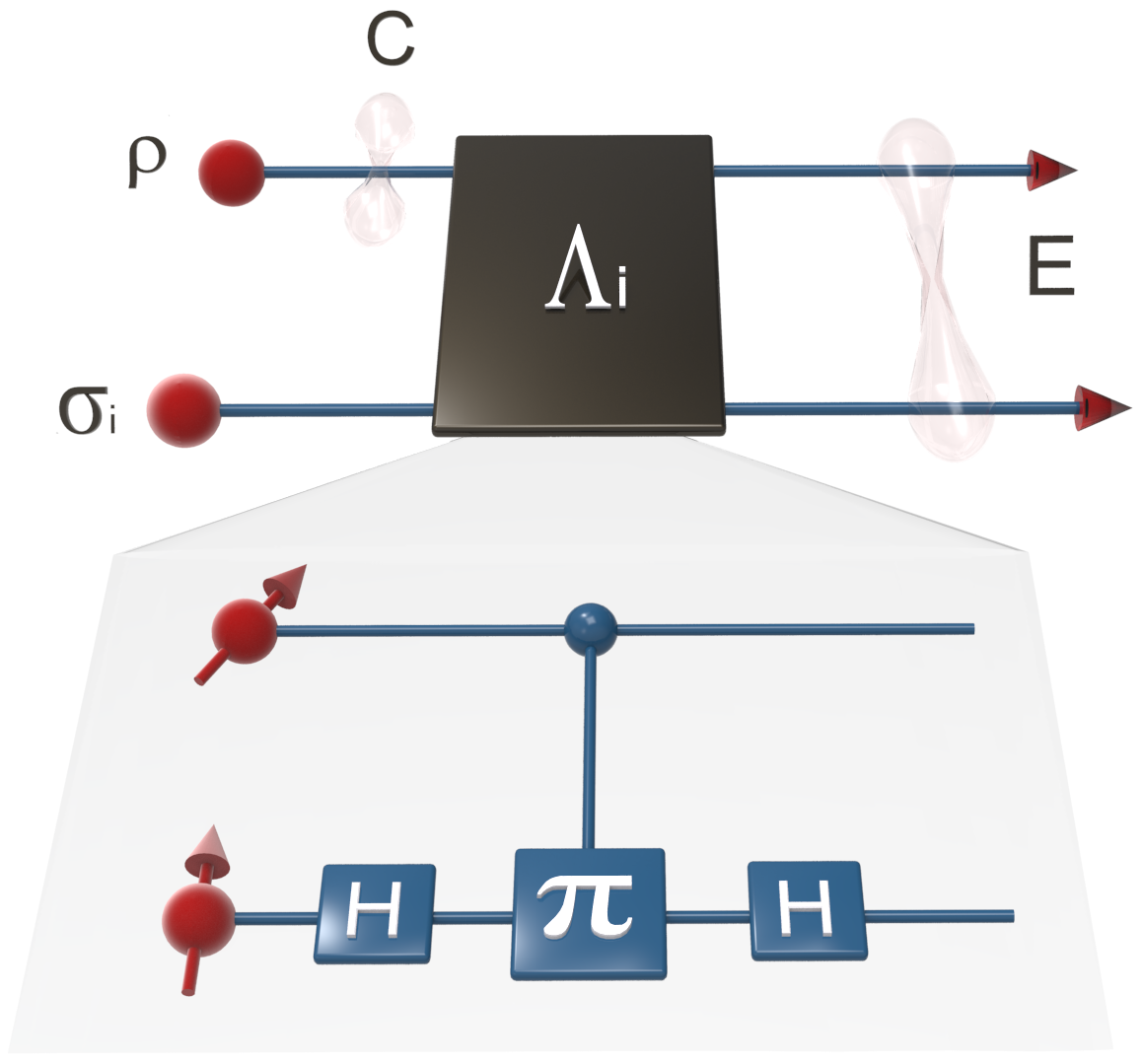} \caption{\textbf{\label{fig:conceptual}The conceptual graph of the conversion
process.} Two individual quantum states are generated and labeled
as the system state and the ancilla state, respectively. The system
state $\rho$ is prepared with a nonzero amount of coherence $C(\rho)$,
while the ancilla is initialized in an incoherent state $\sigma_{i}$
. After an incoherent operation $\Lambda_{i}$ acting on the system
and ancilla, the two-qubit state is either entangled or separable,
depending on whether the initial system state $\rho$ has coherence
or not. Here, we choose the CNOT gate as the optimal incoherent operation,
which is decomposed into one controlled phase gate and two Hadamard
gates.}
\end{figure}

Entanglement activation from coherence has been first studied in \cite{Streltsov2015}.
There, it was shown that any nonzero amount of coherence in a quantum
state $\rho$ can be activated into entanglement by coupling the state
to an incoherent ancilla $\sigma_{i}$ and performing a bipartite
incoherent operation on the total state $\rho\otimes\sigma_{i}$.
On a quantitative level, the amount of coherence in a state $\rho$
bounds the amount of activated entanglement as~\cite{Streltsov2015}
\begin{equation}
E(\Lambda_{i}[\rho\otimes\sigma_{i}])\leq C(\rho),\label{eq:EntCoh-1}
\end{equation}
where $\Lambda_{i}$ is an incoherent operation, and $E$ and $C$
are general distance-based entanglement and coherence monotones, see
Methods section for rigorous definitions and more details. In many
relevant cases, the optimal incoherent operation saturating the inequality
(\ref{eq:EntCoh-1}) is the CNOT gate (see Fig.~\ref{fig:conceptual}).

We will now study this relation from a very general perspective, by
resorting to the resource theory of superposition \cite{Killoran2016,Theurer2017}.
In this theory, the free states $\{\ket{c_{i}}\}$ are not necessarily
mutually orthogonal. Thus, the theory of superposition is more general
than the resource theory of coherence, and is indeed powerful enough
to cover also the resource theory of entanglement, which is obtained
by allowing for continuous sets of free states. Any convex combination
of the free states $\{\ket{c_{i}}\}$ is also a free state, which
is a very natural assumption in any quantum resource theory. Free
operations and further properties of the resource theory of superposition
have been discussed in \cite{Killoran2016,Theurer2017}.

In the following we will study the resource theory of superposition
for a two-qubit system. We assume that each of the qubits has two
pure free states which we denote by $\ket{c_{0}}$ and $\ket{c_{1}}$,
assuming that $0<|\!\braket{c_{0}|c_{1}}\!|<1$. Pure free states
of both qubits have the form $\ket{c_{i}}\otimes\ket{c_{j}}$, and
convex combinations of such states are also free. We will now consider
unitary operations which do not create superpositions of the free
states on both qubits. Following the notion of Ref. \cite{Theurer2017},
we will call these unitaries \emph{superposition-free}. In general,
these unitaries induce the transformation 
\begin{equation}
U\ket{c_{k}}\ket{c_{l}}=e^{i\phi_{kl}}\ket{c_{m}}\ket{c_{n}}\label{eq:U}
\end{equation}
with some phases $e^{i\phi_{kl}}$. Our main question in this context
is the following: \emph{can a bipartite superposition-free unitary
create entanglement?} The answer to this question is affirmative in
the traditional framework of quantum coherence, i.e., for orthogonal
free states $\ket{c_{0}}$ and $\ket{c_{1}}$ \cite{Streltsov2015}.
In this case, the CNOT gate is a superposition-free unitary which
can create entanglement. It is reasonable to believe that these ideas
transfer to the more general concept of superpositions, and that superposition-free
unitaries also allow to create entanglement.

Quite surprisingly, we will see in the following that this is not
the case for the framework considered here. This is the statement
of the following theorem. 
\begin{thm}
\label{thm:1}It is not possible to create entanglement via superposition-free
unitaries on two qubits. 
\end{thm}
\noindent We note that the theorem applies for the case where each
of the qubits has two superposition-free states $\ket{c_{0}}$ and
$\ket{c_{1}}$ with $0<|\!\braket{c_{0}|c_{1}}\!|<1$. The proof of
this theorem will be a combination of several results, which we will
present below.

Before we prove the above theorem, we will first show that every superposition-free
unitary on two qubits can be decomposed into two elementary operations,
which we will denote by $V$ and $W$. The first elementary operation
is the swap gate $V=\sum_{i,j}\ket{ij}\!\bra{ji}$, which corresponds
to an exchange of the two qubits: 
\begin{equation}
V\ket{c_{k}}\ket{c_{l}}\rightarrow\ket{c_{l}}\ket{c_{k}}.\label{eq:V}
\end{equation}
The second elementary operation transforms an initial superposition-free
state $\ket{c_{k}}\ket{c_{l}}$ as follows: 
\begin{equation}
W\ket{c_{k}}\ket{c_{l}}=e^{i\varphi_{k}}\ket{c_{\mathrm{mod}(k+1,2)}}\ket{c_{l}},\label{eq:W}
\end{equation}
where the phases $e^{i\varphi_{k}}$ are defined as 
\begin{equation}
e^{i\varphi_{0}}=1,\,\,\,\,\,\,e^{i\varphi_{1}}=\frac{\braket{c_{0}|c_{1}}}{\braket{c_{1}|c_{0}}}.\label{eq:phases}
\end{equation}
The existence of such a unitary is guaranteed by Lemma~3 in \cite{Marvian2013}
(see also \cite{Chefles2004,Killoran2016}). Note that Eq.~(\ref{eq:W})
defines the action of $W$ onto any pure two-qubit state $\ket{\psi}$,
since any such state can be written as $\ket{\psi}=\sum_{k,l}a_{kl}\ket{c_{k}}\ket{c_{l}}$
with complex numbers $a_{kl}$. Moreover, $W$ can be chosen to be
a local unitary, acting on the first qubit only. With these tools,
we are now in position to prove the following theorem. 
\begin{thm}
\label{thm:2}There exist only eight superposition-free unitaries
for two qubits, which can all be expressed as combinations of $V$
and $W$. 
\end{thm}
\noindent This theorem applies to the same framework of superposition
as Theorem~\ref{thm:1}, i.e., it holds if each qubit has two superposition-free
states $\ket{c_{0}}$ and $\ket{c_{1}}$ with $0<|\!\braket{c_{0}|c_{1}}\!|<1$.
The proof of the theorem is given in Appendix~\ref{sec:Superposition-free-unitaries}.
We list all eight possible transformations in Table~\ref{tab:1}.
\begin{table}
\begin{centering}
\begin{tabular}{ccccccccc}
\hline 
Unitary  & $V^{2}$  & $V$  & $WVW$  & $(VW)^{2}$  & $W$  & $WV$  & $VWV$  & $VW$\tabularnewline
\hline 
$e^{i\phi_{00}}$  & $1$  & $1$  & $1$  & $1$  & $1$  & $1$  & $1$  & $1$\tabularnewline
\hline 
$e^{i\phi_{11}}$  & $1$  & $1$  & $\frac{\braket{c_{0}|c_{1}}^{2}}{\braket{c_{1}|c_{0}}^{2}}$  & $\frac{\braket{c_{0}|c_{1}}^{2}}{\braket{c_{1}|c_{0}}^{2}}$  & $\frac{\braket{c_{0}|c_{1}}}{\braket{c_{1}|c_{0}}}$  & $\frac{\braket{c_{0}|c_{1}}}{\braket{c_{1}|c_{0}}}$  & $\frac{\braket{c_{0}|c_{1}}}{\braket{c_{1}|c_{0}}}$  & $\frac{\braket{c_{0}|c_{1}}}{\braket{c_{1}|c_{0}}}$\tabularnewline
\hline 
$e^{i\phi_{01}}$  & $1$  & $1$  & $\frac{\braket{c_{0}|c_{1}}}{\braket{c_{1}|c_{0}}}$  & $\frac{\braket{c_{0}|c_{1}}}{\braket{c_{1}|c_{0}}}$  & $1$  & $\frac{\braket{c_{0}|c_{1}}}{\braket{c_{1}|c_{0}}}$  & $\frac{\braket{c_{0}|c_{1}}}{\braket{c_{1}|c_{0}}}$  & $1$\tabularnewline
\hline 
$e^{i\phi_{10}}$  & $1$  & $1$  & $\frac{\braket{c_{0}|c_{1}}}{\braket{c_{1}|c_{0}}}$  & $\frac{\braket{c_{0}|c_{1}}}{\braket{c_{1}|c_{0}}}$  & $\frac{\braket{c_{0}|c_{1}}}{\braket{c_{1}|c_{0}}}$  & $1$  & $1$  & $\frac{\braket{c_{0}|c_{1}}}{\braket{c_{1}|c_{0}}}$\tabularnewline
\hline 
\end{tabular}
\par\end{centering}

\caption{\label{tab:1}\textbf{All superposition-free unitaries on two qubits.}
Any superposition-free unitary on two qubits can be expressed as a
product of elementary unitaries $V$ and $W$ given in the main text.
The phases $e^{i\phi_{kl}}$ in the table correspond to the phases
in Eq.~(\ref{eq:U}).}
\end{table}

The tools provided so far give important insight on the structure
of superposition-free unitaries for two qubits and allow us to complete
the proof of Theorem~\ref{thm:1}. For this, it is enough to show
that both elementary operations $V$ and $W$ cannot create entanglement.
Clearly, entanglement cannot be created with the swap unitary $V$.
The second elementary operation $W$ also cannot create entanglement,
as it can be implemented as a local unitary acting on the first qubit only.

At this point it is interesting to compare our results to results
reported in \cite{Killoran2016,Regula1704.04153}. Applied to the setting considered
here, the results of \cite{Killoran2016} imply that superposition
can be converted into entanglement in a \emph{universal} way: there
exists a (not necessarily superposition-free) quantum operation $\Lambda$
which universally converts any state of the form $\ket{\psi}=(\alpha_{0}\ket{c_{0}}+\alpha_{1}\ket{c_{1}})\otimes\ket{c_{0}}$
into an entangled state whenever both coefficients $\alpha_{0}$ and
$\alpha_{1}$ are nonzero. Note that this is not a contradiction to
our results presented above, as the quantum operation $\Lambda$ in
this conversion is not necessarily superposition-free.

We will now show how recent advances in coherence theory can be used
to solve important open questions in the theory of entanglement. For
this, we recall that Eq.~(\ref{eq:EntCoh-1}) also applies to entanglement
and coherence quantifiers based on the trace norm: 
\begin{align}
C_{\mathrm{t}}(\rho) & =\min_{\sigma\in\mathcal{I}}||\rho-\sigma||_{1},\label{eq:Ct}\\
E_{\mathrm{t}}(\rho) & =\min_{\sigma\in\mathcal{S}}||\rho-\sigma||_{1},\label{eq:Et}
\end{align}
where $\mathcal{I}$ and $\mathcal{S}$ are the sets of incoherent
and separable states, respectively. The trace norm $||M||_{1}=\mathrm{Tr}\sqrt{M^{\dagger}M}$
is one of the most important quantities in quantum information theory.
Its significance comes from its operational interpretation, as $p=1/2+||\rho-\sigma||_{1}/4$
is the optimal probability for distinguishing two quantum states $\rho$
and $\sigma$ via quantum measurements. The coherence and entanglement
quantifiers (\ref{eq:Ct}) and (\ref{eq:Et}) thus have the operational
interpretation via the probability to distinguish a state $\rho$
from the set of incoherent and separable states, respectively.

Despite its clear operational significance, it is only very recently
that the trace norm has been investigated within the resource theory
of quantum coherence \cite{Rana2016,Chen2016,Yu2016}, and surprisingly
little is known about the trace norm entanglement $E_{\mathrm{t}}$
\cite{Eisert2006}. Remarkably, it was shown in \cite{Yu2016} that
$C_{\mathrm{t}}$ violates strong monotonicity: the trace norm coherence
of a state can increase on average under a suitable incoherent operation.
We refer to the Methods section for a rigorous definition of strong
monotonicity. As we show in the following theorem, these results also
extend to the trace norm entanglement, thus settling an important
question in entanglement theory which was open for decades. 
\begin{thm}
\label{thm:trace-norm}Trace norm entanglement is not a strong entanglement
monotone. 
\end{thm}
\noindent The proof of the theorem can be found in Appendix~\ref{sec:trace-norm},
where we in fact show that the trace norm entanglement can increase
on average under a local measurement. This finishes the theoretical
part of this work, and we will now focus on experimental entanglement
activation from coherence.

\section*{Experimental Entanglement Activation \protect \\
from Coherence }

\begin{figure*}
\includegraphics[width=2.05\columnwidth]{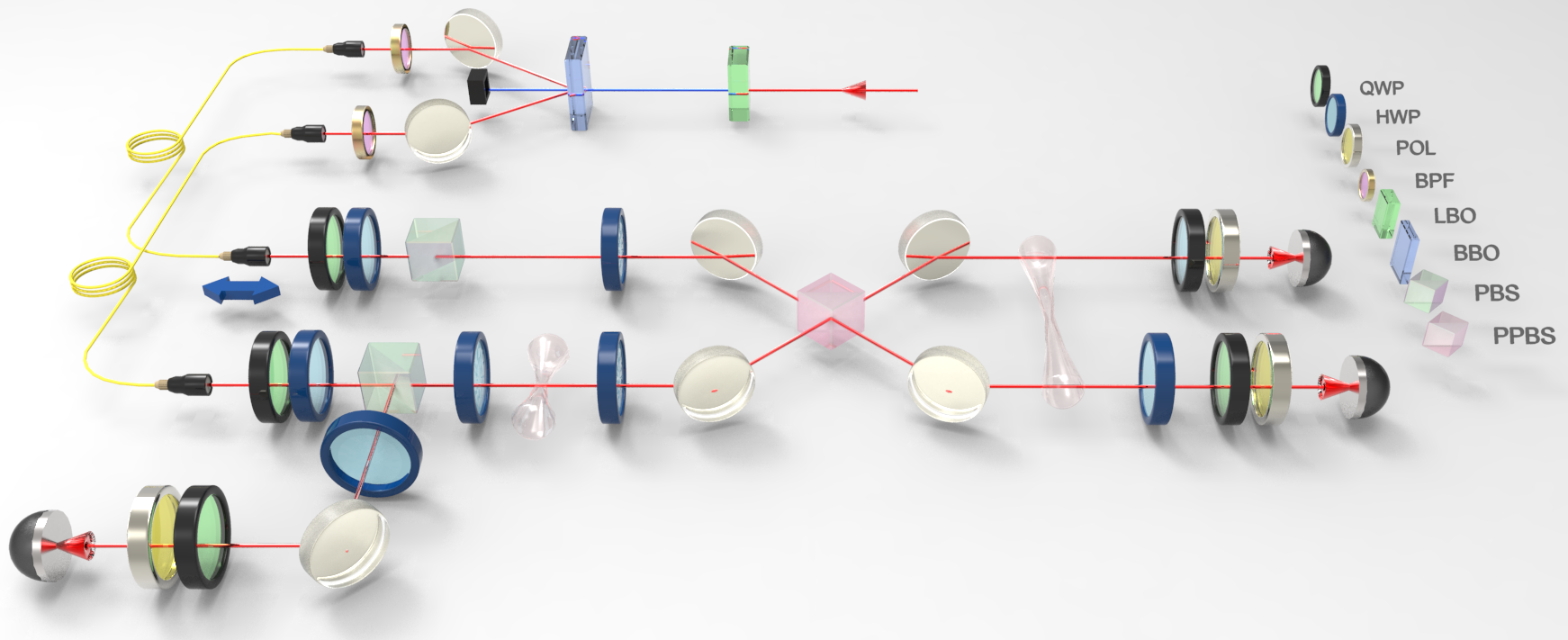} \caption{\textbf{\label{fig:setup}Experimental setup.} Pairs of identical
photons are generated via type-II spontaneous parametric down-conversion
process in a BBO crystal by a 390nm UV laser up-converted from a mode
lock Ti:sapphire oscillator. After passing the $3nm$ band pass filter
(BPF), the photon pairs are coupled into the single mode fibers and
launched to the incoherent operation section. A quarter wave plate
(QWP) and a half wave plate (HWP) are used for polarization compensation.
A combination of HWPs and a partial polarizing beam splitter (PPBS)
acts as the incoherent operation. The system states are prepared with
different amount of coherence by rotating a HWP following
of the PBS. The two-qubit states and an additional copy of the system
states are analyzed by quantum state tomography.}
\end{figure*}

The results presented above impose strong constraints on the possible
activation of superpositions into entanglement. On the other hand,
it is known that activation of entanglement from \emph{coherence}
is possible \cite{Streltsov2015}, i.e., the aforementioned constraints
can be circumvented if the free states $\ket{c_{0}}$ and $\ket{c_{1}}$
are orthogonal. In this case, as is shown in Fig.~\ref{fig:conceptual}, any nonzero amount of coherence in a
state $\rho$ can be converted into entanglement by adding an incoherent
ancilla $\sigma_{i}$ and performing a bipartite incoherent unitary on
the total state $\rho\otimes\sigma_{i}$. As we will see in the following,
such an activation can indeed be performed with current experimental
techniques.

Following our previous discussion, the individual systems will be
qubits. As a quantifier of coherence we will use the $\ell_{1}$-norm
of coherence, which is a strong coherence monotone,
and corresponds to the sum of the absolute values of the off-diagonal
elements \cite{Baumgratz2014}: 
\begin{equation}
C(\rho)=\sum\limits _{i\neq j}\left|\rho_{ij}\right|.\label{eq:coherence}
\end{equation}
After performing a bipartite incoherent operation on the total state
$\rho\otimes\sigma_{i}$, the amount of entanglement in the total
state will be quantified via concurrence $E$. Concurrence is a natural
entanglement quantifier for two-qubit states, as it admits the following
closed expression \cite{Wootters1998}: 
\begin{equation}
E(\rho)=\max\left\{ 0,\lambda_{1}-\lambda_{2}-\lambda_{3}-\lambda_{4}\right\} ,\label{eq:concurrence}
\end{equation}
where $\lambda_{i}$ are the square roots of the eigenvalues of $\rho\tilde{\rho}$
in decreasing order, and $\tilde{\rho}$ is defined as $\tilde{\rho}=(\sigma_{y}\otimes\sigma_{y})\rho^{*}(\sigma_{y}\otimes\sigma_{y})$
with Pauli $y$-matrix $\sigma_{y}$, and complex conjugation is taken
in the computational basis.

\begin{table}[hb!]
\centering{}\begin{tabular}{p{1.2cm}|p{1.2cm}|p{1.2cm}|p{1.2cm}|p{1.2cm}}   \hline\noalign{\smallskip}   \hline\noalign{\smallskip}    ZZ & $\langle{00}\arrowvert$ & $\langle{01}\arrowvert$ & $\langle{10}\arrowvert$ & $\langle{11}\arrowvert$\\   \noalign{\smallskip}\hline\noalign{\smallskip}   $\arrowvert{00}\rangle$ &0.929	&0.034	&0.033	&0.004 \\   $\arrowvert{01}\rangle$	&0.053	&0.914	&0.002	&0.031	 \\   $\arrowvert{10}\rangle$ &0.004 &0.002 &0.159	&0.835	\\   $\arrowvert{11}\rangle$ &0.001 &0.005 &0.816	&0.178	\\   \noalign{\smallskip}\hline   \end{tabular}   \begin{tabular}{p{1.2cm}|p{1.2cm}|p{1.2cm}|p{1.2cm}|p{1.2cm}}   \hline\noalign{\smallskip}    XX & $\langle{00}\arrowvert$ & $\langle{01}\arrowvert$ & $\langle{10}\arrowvert$ & $\langle{11}\arrowvert$\\   \noalign{\smallskip}\hline\noalign{\smallskip}   $\arrowvert{00}\rangle$ &0.896	&0.004	&0.099	&0.001 \\   $\arrowvert{01}\rangle$	&0.002	&0.173	&0.001	&0.824	 \\   $\arrowvert{10}\rangle$ &0.103 &0.002 &0.892	&0.003	\\   $\arrowvert{11}\rangle$ &0.001 &0.827 &0.001	&0.171	\\   \noalign{\smallskip}\hline   \hline\noalign{\smallskip}   \end{tabular}  
\caption{\label{tab:truthtable}Truth table of the CNOT gate.}
\end{table}

\begin{figure*}
\includegraphics[width=2.05\columnwidth]{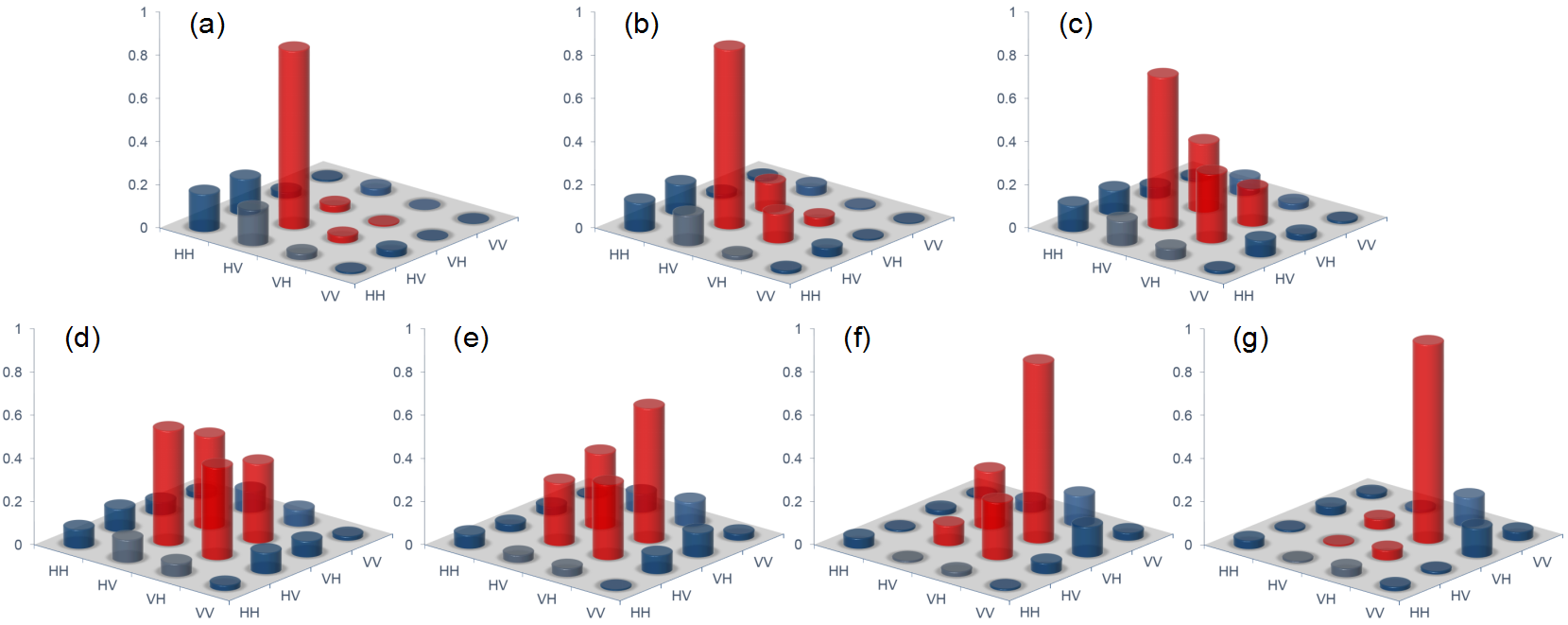} \caption{\textbf{\label{fig:tomography}Experimental results of two-qubit tomography.}
The density matrices with different system states $cos(\vartheta)\arrowvert{H}\rangle+sin(\vartheta)\arrowvert{V}\rangle$
as the input state by scanning different polarizations (a)$\vartheta=0^{\circ}$;
(b)$\vartheta=15^{\circ}$; (c)$\vartheta=30^{\circ}$; (d)$\vartheta=45^{\circ}$;
(e)$\vartheta=60^{\circ}$; (f)$\vartheta=75^{\circ}$; (g)$\vartheta=90^{\circ}$.
From (a) to (d), it is obvious that the generated entangled states
vary from separable states to maximal entangled state while from (e)
to (g), the entanglement gradually decreases due to the decline of
the coherence.}
\end{figure*}

As we show in Appendix~\ref{sec:concurrence}, Eq.~(\ref{eq:EntCoh-1})
also applies in this situation, i.e., the amount of coherence in the
state $\rho$ bounds the amount of concurrence that can be activated
from the state via incoherent operations. Moreover, the optimal incoherent
operation in the above setting is the CNOT gate, as it allows to saturate
the inequality~(\ref{eq:EntCoh-1}). We also note that for the systems considered here the $\ell_{1}$-norm coherence
coincides with the trace norm coherence~\cite{Shao2015}. Thus, the
results discussed in this section also hold if $C$ is the trace norm
coherence defined in Eq.~(\ref{eq:Ct}).

Here, we experimentally verify this relation between coherence and
entanglement by the means of quantum optics, using the fact that polarization
is easy to manipulate with high precision. By utilizing the phase
flip introduced by second order interference, we construct the incoherent
operation with a combination of a controlled phase gate and two Hadamard
gates. We prepare a set of system states with different amount of
coherence, and observe that coherence and entanglement
are highly correlated with acceptable errors under the state of art
of optical CNOT operation \cite{Kiesel2005,Okamoto2005,Crespi2011}.

The sketch of our experiment setup is shown in Fig.~\ref{fig:setup}.
It can be divided into three parts: the preparation of identical photons,
the incoherent operation and the state analysis module. We use a mode
lock Ti:sapphire oscillator emitting $130fs$ pulses centered at
$780nm$ with a repetition rate of $77MHz$. The near-infrared light
is frequency doubled to ultraviolet light of $390nm$ in a $1.3mm$
thick $LiB_{3}O_{5}$ (LBO) crystal. Two identical
photons are created by pumping a $2mm$ thick $\beta-BaB_{2}O_{4}$
(BBO) crystal via a type-II spontaneous parametric
down-conversion process in a beam-like scheme \cite{Kim2003,Kwiat1995}.
Two $3nm$ band pass filters are used to improve
the visibility of interference for it ensures the spectral indistinguishablity
of the photon pairs. The photons are coupled into the single mode
fibers, with one serving as the system photon while the other one
as the ancilla photon. A quarter wave plate and a
half wave plate are used in both arms to compensate the polarization
rotation induced by the single mode fibers.

The two indistinguishable photons are then injected into the CNOT
gate module based on the second-order interference \cite{Hong1987}.
The key feature in this optical CNOT gate scheme is a partial polarizing
beam splitter (PPBS), which perfectly reflects vertical polarization
and reflects (transmits) 1/3 (2/3) of horizontal polarization. We
mount the coupler for the ancilla photon on a one-dimensional translation
stage to ensure the temporal overlap between the photon pairs. The
ideal HOM interference visibility on this PPBS is $V_{th}=80\%$ and
we experimentally achieve $V_{exp}=67.9\pm1.0\%$. The
relative visibility is $V_{re}=V_{exp}/V_{th}=84.9\%$. The mismatch
can be attributed to the imperfection of the PPBS, whose reflection
ratio of the horizontal polarization $29\%$ deviates from the ideal
value of 33.3. In order to evaluate the performance of the CNOT
gate, we measure the truth tables and estimate the process fidelity
\cite{Hofmann2005}. In the $ZZ$ basis, we define the computational
basis as $\arrowvert{0}\rangle_{z}=\arrowvert{H}\rangle$ and $\arrowvert{1}\rangle_{z}=\arrowvert{V}\rangle$
for the control qubit and $\arrowvert{0}\rangle_{z}=\arrowvert{D}\rangle$
and $\arrowvert{1}\rangle_{z}=\arrowvert{A}\rangle$ for the target
qubit. The CNOT gate flips the target qubit when the control qubit
is $\arrowvert{1}\rangle_{z}$. In the $XX$ basis, it is equivalent
to transform the bases using a Hardamard gate, where the control qubit
is encoded in $\arrowvert{D}\rangle-\arrowvert{A}\rangle$ basis and
the target qubit in $\arrowvert{H}\rangle-\arrowvert{V}\rangle$ basis.
Table~\ref{tab:truthtable} gives the normalized possibilities of
all the combinations with four different input and output states in
both $ZZ$ and $XX$ basis. We can see that the control and the target
qubit swap in the $XX$ basis, where the control qubit remains unchanged
when the target qubit is $\arrowvert{0}\rangle_{x}$ and flips when
the target qubit is $\arrowvert{1}\rangle_{x}$.

\begin{figure}
\includegraphics[width=1\columnwidth]{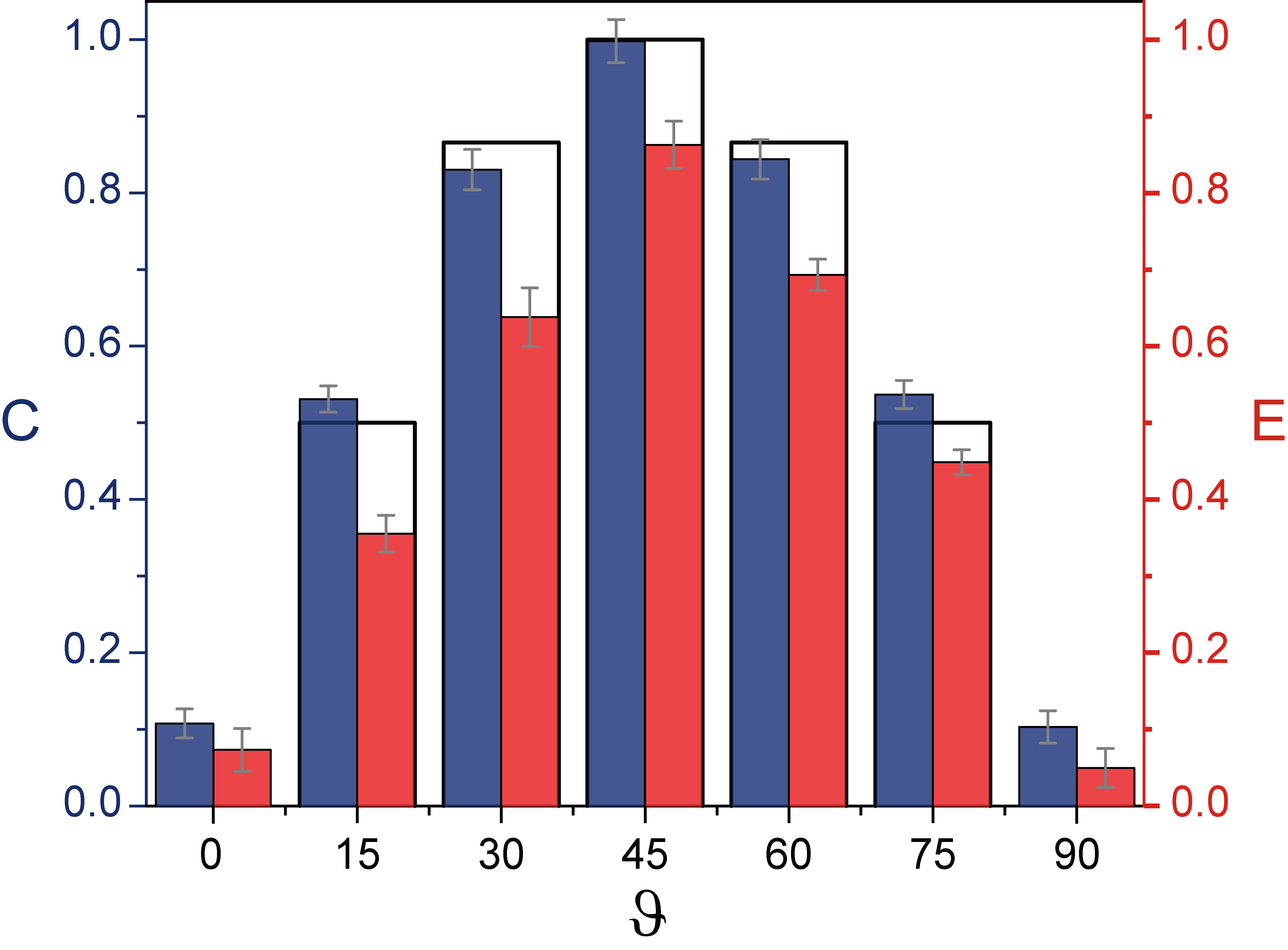} \caption{\textbf{\label{fig:equivalence}Activation of entanglement from coherence.}
The blue bars represent the measured coherence of system qubit as
quantified by the $\ell_{1}$-norm of coherence in Eq.~(\ref{eq:coherence}).
The red bars represent the measured entanglement in the two-qubit
state after the incoherent operation, quantified by the concurrence
in Eq.~(\ref{eq:concurrence}). The outside frames are the theoretical
prediction for coherence and entanglement. The experimental results
show the same tendency as we vary the parameter $\vartheta$. All
error bars are estimated by the Monte Carlo simulation with 1000 rounds
by assuming the Poissonian distribution of the photon statistics.
}
\end{figure}

The fidelity can be defined as the average value of the possibility
to get the correct output over all inputs. From this definition we
can calculate $F_{zz}=0.87$ and $F_{xx}=0.86$. These two complementary
fidelity values can bound the quantum process fidelity according
to \cite{Hofmann2005} 
\begin{equation}
F_{zz}+F_{xx}-1\le F_{process}\le Min\{F_{zz},F_{xx}\}.
\end{equation}
Thus, we can estimate $0.73\le F_{process}\le0.86$. The process fidelity
also benchmarks the minimal entanglement capability
$C\ge2F_{process}-1$, as in our case, the result is larger than~$0.47$.

After experimentally characterizing the incoherent operation, we generate
a series of quantum states: 
\begin{equation}
\begin{aligned}\rho=cos^{2}(\vartheta)\arrowvert{H}\rangle\langle{H}\arrowvert+cos(\vartheta)sin(\vartheta)\arrowvert{H}\rangle\langle{V}\arrowvert\\
+sin(\vartheta)cos(\vartheta)\arrowvert{V}\rangle\langle{H}\arrowvert+sin^{2}(\vartheta)\arrowvert{V}\rangle\langle{V}\arrowvert
\end{aligned}
\end{equation}
By choosing different polarization parameter $\vartheta$, we are
able to tune the corresponding amount of coherence in the system qubit in $\{ \arrowvert{H}\rangle,\arrowvert{V}\rangle \}$ basis.
We split the system qubit on a beam splitter and prepare the two copies
with the same polarization to test the relationship between coherence
and entanglement. The ancilla qubit is fixed to $\sigma_{i}=\arrowvert{H}\rangle\langle{H}\arrowvert$
as an incoherent state during the whole experiment. We first conduct
the one-qubit tomography with a combination of quarter
wave plate and polarizer to reconstruct the $2\times2$ density matrix
of the system qubit \cite{James2001} and further estimate the amount
of coherence defined in Eq.~(\ref{eq:coherence}). The other copy
of the system qubit is guided to the CNOT gate and interferes with
the ancilla qubit on the PPBS. After the incoherent operation, the
two-qubit tomography is used to evaluate the entanglement, as quantified
via concurrence in Eq.~(\ref{eq:concurrence}).

In our experiment, we prepare seven different system states to test
the relation between coherence and entanglement in Eq.~(\ref{eq:EntCoh-1}).
As we vary the coherence parameter, the density matrix of the entanglement
states generated by the incoherent operation correspondingly alter,
as demonstrated in Fig.~\ref{fig:tomography}, from separable states
to maximal entangled state. To further evaluate the relation between
coherence and entanglement, we compare their exact values in Fig.~\ref{fig:equivalence}.
The blue bars represent the amount of coherence and the red bars represent
the amount of entanglement. The outside frames are the theoretical
prediction by considering the ideal cases.

With high-extinction polarization device, we are able
to prepare the maximal coherence state $\arrowvert{D}\rangle=(\arrowvert{H}\rangle+\arrowvert{V}\rangle)/\sqrt{2}$
and the measured coherence is up to $C=0.999$, which
is very close to ideal scenario. The measured entanglement of the
generated entangled state is $E=0.864$. In the next step we decrease
the coherence of the system qubit and the corresponding entanglement
changes with the same tendency. The system with the
minimal coherence in our experiment has $C=0.09$, and the corresponding
activated entanglement between the two qubit is measured to be $E=0.07$.
Given the imperfection of the incoherent operation, certain mismatch
exists between the measured entanglement and coherence. A considerably
high conversion efficiency can be expected after certain optimization
of the device.

\section*{Conclusions}

In this work we explored the possibilities and limitation to activate
entanglement from quantum coherence and superposition. While coherence
can be activated into entanglement via free unitaries of the theory
\cite{Streltsov2015}, we have shown that such an activation is not
possible within a more general theory of quantum superposition. We
have rigorously proven this statement for a general two-qubit system,
where each of the qubits has two superposition-free states $\ket{c_{0}}$
and $\ket{c_{1}}$ with $0<|\braket{c_{0}|c_{1}}|<1$. We have further
shown that only eight superposition-free unitaries are possible in
this setting, and all of them can be represented in terms of two elementary
operations.

An important consequence of our discussion is the finding that in
the general framework of superposition considered here there is no
unitary which corresponds to the action of a CNOT gate, i.e., which
flips the state of the second qubit between $\ket{c_{0}}$ and $\ket{c_{1}}$
conditioned on the first qubit being in the state $\ket{c_{0}}$ or
$\ket{c_{1}}$. Such a CNOT gate exists only in the more restricted
resource theory of coherence, which arises in our framework in the
limit of orthogonal states $\ket{c_{0}}$ and $\ket{c_{1}}$. These
results are analogous to the no-cloning theorem \cite{Wootters1982},
i.e., while it is not possible to clone a general
quantum state, cloning is possible in a more restricted theory, where
the considered states are mutually orthogonal.

We have experimentally demonstrated that entanglement
activation from coherence is indeed possible. We have prepared single-qubit
states with different values of coherence by using polarized photons
and experimentally activated coherence into entanglement via an optical
CNOT gate which is the optimal incoherent operation in the considered
setting. We have then compared the amount of final entanglement to
the amount of initial coherence, finding a good agreement between
theory and experiment. Both quantities clearly show the same tendency:
a large amount of initial coherence leads to a large amount of activated
entanglement.

We also note that related results have been presented
very recently in \cite{Wu1710.01738}, where cyclic interconversion
between coherence and entanglement has been demonstrated experimentally,
based on the framework of assisted coherence distillation \cite{Chitambar2016,Streltsov2017}
and coherence activation from entanglement \cite{Streltsov2015} and
quantum discord~\cite{Modi2012,Ma2016}.

Our work also lead to a surprising result in entanglement
theory, showing that the trace norm entanglement violates strong monotonicity.
This solves an important question in quantum information theory which
was open for decades, and clearly demonstrate how recent developments
on the resource theory of quantum coherence \cite{Streltsov2016}
can be applied for advancing other research areas
of quantum information and technology.

\section*{Methods}

An important question in any quantum resource theory is to quantify
the amount of the resource in a given quantum state. A general resource
quantifier $\mathcal{R}$ should at least have the following property:
\begin{equation}
\mathcal{R}(\Lambda_{f}[\rho])\leq\mathcal{R}(\rho),\label{eq:monotone}
\end{equation}
where $\Lambda_{f}$ is a free operation of the resource theory. In
entanglement theory, $\Lambda_{f}$ are usually chosen to be \emph{local
operations and classical communication} \cite{Horodecki2009}. In
the resource theory of coherence, a possible choice for $\Lambda_{f}$
are \emph{incoherent operations} introduced in \cite{Baumgratz2014},
and alternative frameworks have also been discussed recently \cite{Winter2016,Yadin2016},
see also the review \cite{Streltsov2016} and references therein.

Any nonnegative function $\mathcal{R}$ which fulfills Eq.~(\ref{eq:monotone})
is called\emph{ monotone} of the corresponding resource theory. A
very general family are distance-based monotones 
\begin{equation}
\mathcal{R}_{D}(\rho)=\inf_{\sigma\in\mathcal{F}}D(\rho,\sigma),
\end{equation}
where $\mathcal{F}$ is the set of free states and $D$ is a suitable
distance. The quantity $\mathcal{R}_{D}$ fulfills monotonicity~(\ref{eq:monotone})
for any distance $D$ which is contractive under quantum operations:
$D(\Lambda[\rho],\Lambda[\sigma])\leq D(\rho,\sigma)$. Important
examples for such distances are the quantum relative entropy $S(\rho||\sigma)=\mathrm{Tr}[\rho\log_{2}\rho]-\mathrm{Tr}[\rho\log_{2}\sigma]$
and the trace distance $D_{\mathrm{t}}(\rho,\sigma)=\frac{1}{2}||\rho-\sigma||_{1}$
with the trace norm $||M||_{1}=\mathrm{Tr}\sqrt{M^{\dagger}M}$.

In many resource theories it is also important to consider \emph{selective
free operations}. Here, an initial quantum state $\rho$ is transformed
into an ensemble 
\begin{equation}
\rho\rightarrow\{q_{i},\sigma_{i}\}\label{eq:selective}
\end{equation}
with probabilities $q_{i}$ and quantum states $\sigma_{i}$. In entanglement
theory, this is motivated by the fact that the parties can -- in principle
-- record the outcome of their local measurements. Each state $\sigma_{i}$
then corresponds to the state of the system for a particular sequence
of local measurement outcomes, with a corresponding overall probability
$q_{i}$. A similar approach has been taken recently in the resource
theory of coherence \cite{Baumgratz2014,Winter2016,Streltsov2016,Yadin2016}.

For a resource theory with selective free operations as given in Eq.~(\ref{eq:selective}),
it is reasonable to demand that the corresponding resource quantifier
$\mathcal{R}$ admits \emph{strong monotonicity}: 
\begin{equation}
\sum_{i}q_{i}\mathcal{R}(\sigma_{i})\leq\mathcal{R}(\rho)\label{eq:strong-monotone}
\end{equation}
for any ensemble $\{q_{i},\sigma_{i}\}$ which can be obtained from
the state $\rho$ by the means of selective free operations. The motivation
for this requirement is similar to the standard monotonicity (\ref{eq:monotone}):
the resource should not increase on average even if the outcomes of
free measurements are recorded. Entanglement and coherence monotones
based on the relative entropy fulfill strong monotonicity \cite{Vedral1997,Baumgratz2014}.
As was shown in \cite{Yu2016}, the trace norm coherence violates
strong monotonicity. As we prove in Appendix~\ref{sec:trace-norm},
strong monotonicity is also violated by the trace norm entanglement.
Note that strong monotonicity~(\ref{eq:strong-monotone}) implies
monotonicity~(\ref{eq:monotone}) if $\mathcal{R}$ is convex.

\section*{Acknowledgements}

We acknowledge discussion with P. Horodecki and J.-W. Pan. This work was supported by National Key R\&D Program of China (2017YFA0303700), National Natural Science Foundation of China (NSFC) (11374211, 11690033, 11275131, 11571313), Shanghai Municipal Education Commission (SMEC)(16SG09, 2017-01-07-00-02-E00049), Science and Technology Commission of Shanghai Municipality (STCSM) (15QA1402200, 16JC1400405) and Open fund from HPCL (201511-01). X.-M.J. acknowledges support from the National Young 1000 Talents Plan. A.S. was supported by the National Science Center in Poland (POLONEZ UMO-2016/21/P/ST2/04054). S.R. and M.L. acknowledge support from EC grants OSYRIS (ERC-2013-ADG No. 339106) and QUIC (H2020-FETPROACT-2014 No. 641122), the Spanish MINECO grants Severo Ochoa (SEV-2015-0522), FISICATEAMO (FIS2016-79508-P), MINECO CLUSTER (ICFO15- EE-3785), the Generalitat de Catalunya (2014 SGR 874 and 5 CERCA/Program) and the Fundaci\'o Privada Cellex. 

 \bibliographystyle{apsrev4-1}

\begin{thebibliography}{46}%
\makeatletter
\providecommand \@ifxundefined [1]{%
 \@ifx{#1\undefined}
}%
\providecommand \@ifnum [1]{%
 \ifnum #1\expandafter \@firstoftwo
 \else \expandafter \@secondoftwo
 \fi
}%
\providecommand \@ifx [1]{%
 \ifx #1\expandafter \@firstoftwo
 \else \expandafter \@secondoftwo
 \fi
}%
\providecommand \natexlab [1]{#1}%
\providecommand \enquote  [1]{``#1''}%
\providecommand \bibnamefont  [1]{#1}%
\providecommand \bibfnamefont [1]{#1}%
\providecommand \citenamefont [1]{#1}%
\providecommand \href@noop [0]{\@secondoftwo}%
\providecommand \href [0]{\begingroup \@sanitize@url \@href}%
\providecommand \@href[1]{\@@startlink{#1}\@@href}%
\providecommand \@@href[1]{\endgroup#1\@@endlink}%
\providecommand \@sanitize@url [0]{\catcode `\\12\catcode `\$12\catcode
  `\&12\catcode `\#12\catcode `\^12\catcode `\_12\catcode `\%12\relax}%
\providecommand \@@startlink[1]{}%
\providecommand \@@endlink[0]{}%
\providecommand \url  [0]{\begingroup\@sanitize@url \@url }%
\providecommand \@url [1]{\endgroup\@href {#1}{\urlprefix }}%
\providecommand \urlprefix  [0]{URL }%
\providecommand \Eprint [0]{\href }%
\providecommand \doibase [0]{http://dx.doi.org/}%
\providecommand \selectlanguage [0]{\@gobble}%
\providecommand \bibinfo  [0]{\@secondoftwo}%
\providecommand \bibfield  [0]{\@secondoftwo}%
\providecommand \translation [1]{[#1]}%
\providecommand \BibitemOpen [0]{}%
\providecommand \bibitemStop [0]{}%
\providecommand \bibitemNoStop [0]{.\EOS\space}%
\providecommand \EOS [0]{\spacefactor3000\relax}%
\providecommand \BibitemShut  [1]{\csname bibitem#1\endcsname}%
\let\auto@bib@innerbib\@empty
\bibitem [{\citenamefont {Schr\"odinger}(1935)}]{Schrodinger1935a}%
  \BibitemOpen
  \bibfield  {author} {\bibinfo {author} {\bibfnamefont {E.}~\bibnamefont
  {Schr\"odinger}},\ }\href {\doibase 10.1007/BF01491891} {\bibfield  {journal}
  {\bibinfo  {journal} {Naturwissenschaften}\ }\textbf {\bibinfo {volume}
  {23}},\ \bibinfo {pages} {807} (\bibinfo {year} {1935})}\BibitemShut
  {NoStop}%
\bibitem [{\citenamefont {Einstein}\ \emph {et~al.}(1935)\citenamefont
  {Einstein}, \citenamefont {Podolsky},\ and\ \citenamefont
  {Rosen}}]{EinsteinPhysRev.47.777}%
  \BibitemOpen
  \bibfield  {author} {\bibinfo {author} {\bibfnamefont {A.}~\bibnamefont
  {Einstein}}, \bibinfo {author} {\bibfnamefont {B.}~\bibnamefont {Podolsky}},
  \ and\ \bibinfo {author} {\bibfnamefont {N.}~\bibnamefont {Rosen}},\ }\href
  {\doibase 10.1103/PhysRev.47.777} {\bibfield  {journal} {\bibinfo  {journal}
  {Phys. Rev.}\ }\textbf {\bibinfo {volume} {47}},\ \bibinfo {pages} {777}
  (\bibinfo {year} {1935})}\BibitemShut {NoStop}%
\bibitem [{\citenamefont {Horodecki}\ \emph {et~al.}(2009)\citenamefont
  {Horodecki}, \citenamefont {Horodecki}, \citenamefont {Horodecki},\ and\
  \citenamefont {Horodecki}}]{Horodecki2009}%
  \BibitemOpen
  \bibfield  {author} {\bibinfo {author} {\bibfnamefont {R.}~\bibnamefont
  {Horodecki}}, \bibinfo {author} {\bibfnamefont {P.}~\bibnamefont
  {Horodecki}}, \bibinfo {author} {\bibfnamefont {M.}~\bibnamefont
  {Horodecki}}, \ and\ \bibinfo {author} {\bibfnamefont {K.}~\bibnamefont
  {Horodecki}},\ }\href {\doibase 10.1103/RevModPhys.81.865} {\bibfield
  {journal} {\bibinfo  {journal} {Rev. Mod. Phys.}\ }\textbf {\bibinfo {volume}
  {81}},\ \bibinfo {pages} {865} (\bibinfo {year} {2009})}\BibitemShut
  {NoStop}%
\bibitem [{\citenamefont {{Streltsov}}\ \emph {et~al.}(2016)\citenamefont
  {{Streltsov}}, \citenamefont {{Adesso}},\ and\ \citenamefont
  {{Plenio}}}]{Streltsov2016}%
  \BibitemOpen
  \bibfield  {author} {\bibinfo {author} {\bibfnamefont {A.}~\bibnamefont
  {{Streltsov}}}, \bibinfo {author} {\bibfnamefont {G.}~\bibnamefont
  {{Adesso}}}, \ and\ \bibinfo {author} {\bibfnamefont {M.~B.}\ \bibnamefont
  {{Plenio}}},\ }\href {https://arxiv.org/abs/1609.02439} {\bibfield  {journal}
  {\bibinfo  {journal} {arXiv:1609.02439}\ } (\bibinfo {year}
  {2016})}\BibitemShut {NoStop}%
\bibitem [{\citenamefont {Vedral}\ \emph {et~al.}(1997)\citenamefont {Vedral},
  \citenamefont {Plenio}, \citenamefont {Rippin},\ and\ \citenamefont
  {Knight}}]{Vedral1997}%
  \BibitemOpen
  \bibfield  {author} {\bibinfo {author} {\bibfnamefont {V.}~\bibnamefont
  {Vedral}}, \bibinfo {author} {\bibfnamefont {M.~B.}\ \bibnamefont {Plenio}},
  \bibinfo {author} {\bibfnamefont {M.~A.}\ \bibnamefont {Rippin}}, \ and\
  \bibinfo {author} {\bibfnamefont {P.~L.}\ \bibnamefont {Knight}},\ }\href
  {\doibase 10.1103/PhysRevLett.78.2275} {\bibfield  {journal} {\bibinfo
  {journal} {Phys. Rev. Lett.}\ }\textbf {\bibinfo {volume} {78}},\ \bibinfo
  {pages} {2275} (\bibinfo {year} {1997})}\BibitemShut {NoStop}%
\bibitem [{\citenamefont {Baumgratz}\ \emph {et~al.}(2014)\citenamefont
  {Baumgratz}, \citenamefont {Cramer},\ and\ \citenamefont
  {Plenio}}]{Baumgratz2014}%
  \BibitemOpen
  \bibfield  {author} {\bibinfo {author} {\bibfnamefont {T.}~\bibnamefont
  {Baumgratz}}, \bibinfo {author} {\bibfnamefont {M.}~\bibnamefont {Cramer}}, \
  and\ \bibinfo {author} {\bibfnamefont {M.~B.}\ \bibnamefont {Plenio}},\
  }\href {\doibase 10.1103/PhysRevLett.113.140401} {\bibfield  {journal}
  {\bibinfo  {journal} {Phys. Rev. Lett.}\ }\textbf {\bibinfo {volume} {113}},\
  \bibinfo {pages} {140401} (\bibinfo {year} {2014})}\BibitemShut {NoStop}%
\bibitem [{\citenamefont {Werner}(1989)}]{WernerPhysRevA.40.4277}%
  \BibitemOpen
  \bibfield  {author} {\bibinfo {author} {\bibfnamefont {R.~F.}\ \bibnamefont
  {Werner}},\ }\href {\doibase 10.1103/PhysRevA.40.4277} {\bibfield  {journal}
  {\bibinfo  {journal} {Phys. Rev. A}\ }\textbf {\bibinfo {volume} {40}},\
  \bibinfo {pages} {4277} (\bibinfo {year} {1989})}\BibitemShut {NoStop}%
\bibitem [{\citenamefont {Bennett}\ \emph {et~al.}(1993)\citenamefont
  {Bennett}, \citenamefont {Brassard}, \citenamefont {Cr\'epeau}, \citenamefont
  {Jozsa}, \citenamefont {Peres},\ and\ \citenamefont
  {Wootters}}]{BennettPhysRevLett.70.1895}%
  \BibitemOpen
  \bibfield  {author} {\bibinfo {author} {\bibfnamefont {C.~H.}\ \bibnamefont
  {Bennett}}, \bibinfo {author} {\bibfnamefont {G.}~\bibnamefont {Brassard}},
  \bibinfo {author} {\bibfnamefont {C.}~\bibnamefont {Cr\'epeau}}, \bibinfo
  {author} {\bibfnamefont {R.}~\bibnamefont {Jozsa}}, \bibinfo {author}
  {\bibfnamefont {A.}~\bibnamefont {Peres}}, \ and\ \bibinfo {author}
  {\bibfnamefont {W.~K.}\ \bibnamefont {Wootters}},\ }\href {\doibase
  10.1103/PhysRevLett.70.1895} {\bibfield  {journal} {\bibinfo  {journal}
  {Phys. Rev. Lett.}\ }\textbf {\bibinfo {volume} {70}},\ \bibinfo {pages}
  {1895} (\bibinfo {year} {1993})}\BibitemShut {NoStop}%
\bibitem [{\citenamefont {Jin}\ \emph {et~al.}(2010)\citenamefont {Jin},
  \citenamefont {Ren}, \citenamefont {Yang}, \citenamefont {Yi}, \citenamefont
  {Zhou}, \citenamefont {Xu}, \citenamefont {Wang}, \citenamefont {Yang},
  \citenamefont {Hu}, \citenamefont {Jiang}, \citenamefont {Yang},
  \citenamefont {Yin}, \citenamefont {Chen}, \citenamefont {Peng},\ and\
  \citenamefont {Pan}}]{Jin2010}%
  \BibitemOpen
  \bibfield  {author} {\bibinfo {author} {\bibfnamefont {X.-M.}\ \bibnamefont
  {Jin}}, \bibinfo {author} {\bibfnamefont {J.-G.}\ \bibnamefont {Ren}},
  \bibinfo {author} {\bibfnamefont {B.}~\bibnamefont {Yang}}, \bibinfo {author}
  {\bibfnamefont {Z.-H.}\ \bibnamefont {Yi}}, \bibinfo {author} {\bibfnamefont
  {F.}~\bibnamefont {Zhou}}, \bibinfo {author} {\bibfnamefont {X.-F.}\
  \bibnamefont {Xu}}, \bibinfo {author} {\bibfnamefont {S.-K.}\ \bibnamefont
  {Wang}}, \bibinfo {author} {\bibfnamefont {D.}~\bibnamefont {Yang}}, \bibinfo
  {author} {\bibfnamefont {Y.-F.}\ \bibnamefont {Hu}}, \bibinfo {author}
  {\bibfnamefont {S.}~\bibnamefont {Jiang}}, \bibinfo {author} {\bibfnamefont
  {T.}~\bibnamefont {Yang}}, \bibinfo {author} {\bibfnamefont {H.}~\bibnamefont
  {Yin}}, \bibinfo {author} {\bibfnamefont {K.}~\bibnamefont {Chen}}, \bibinfo
  {author} {\bibfnamefont {C.-Z.}\ \bibnamefont {Peng}}, \ and\ \bibinfo
  {author} {\bibfnamefont {J.-W.}\ \bibnamefont {Pan}},\ }\href {\doibase
  10.1038/nphoton.2010.87} {\bibfield  {journal} {\bibinfo  {journal} {Nat.
  Photon.}\ }\textbf {\bibinfo {volume} {4}},\ \bibinfo {pages} {376} (\bibinfo
  {year} {2010})}\BibitemShut {NoStop}%
\bibitem [{\citenamefont {Ekert}(1991)}]{EkertPhysRevLett.67.661}%
  \BibitemOpen
  \bibfield  {author} {\bibinfo {author} {\bibfnamefont {A.~K.}\ \bibnamefont
  {Ekert}},\ }\href {\doibase 10.1103/PhysRevLett.67.661} {\bibfield  {journal}
  {\bibinfo  {journal} {Phys. Rev. Lett.}\ }\textbf {\bibinfo {volume} {67}},\
  \bibinfo {pages} {661} (\bibinfo {year} {1991})}\BibitemShut {NoStop}%
\bibitem [{\citenamefont {{Horodecki}}\ \emph {et~al.}(2005)\citenamefont
  {{Horodecki}}, \citenamefont {{Oppenheim}},\ and\ \citenamefont
  {{Winter}}}]{Horodecki2005}%
  \BibitemOpen
  \bibfield  {author} {\bibinfo {author} {\bibfnamefont {M.}~\bibnamefont
  {{Horodecki}}}, \bibinfo {author} {\bibfnamefont {J.}~\bibnamefont
  {{Oppenheim}}}, \ and\ \bibinfo {author} {\bibfnamefont {A.}~\bibnamefont
  {{Winter}}},\ }\href {\doibase 10.1038/nature03909} {\bibfield  {journal}
  {\bibinfo  {journal} {Nature}\ }\textbf {\bibinfo {volume} {436}},\ \bibinfo
  {pages} {673} (\bibinfo {year} {2005})}\BibitemShut {NoStop}%
\bibitem [{\citenamefont {Yin}\ \emph {et~al.}(2017)\citenamefont {Yin},
  \citenamefont {Cao}, \citenamefont {Li}, \citenamefont {Liao}, \citenamefont
  {Zhang}, \citenamefont {Ren}, \citenamefont {Cai}, \citenamefont {Liu},
  \citenamefont {Li}, \citenamefont {Dai}, \citenamefont {Li}, \citenamefont
  {Lu}, \citenamefont {Gong}, \citenamefont {Xu}, \citenamefont {Li},
  \citenamefont {Li}, \citenamefont {Yin}, \citenamefont {Jiang}, \citenamefont
  {Li}, \citenamefont {Jia}, \citenamefont {Ren}, \citenamefont {He},
  \citenamefont {Zhou}, \citenamefont {Zhang}, \citenamefont {Wang},
  \citenamefont {Chang}, \citenamefont {Zhu}, \citenamefont {Liu},
  \citenamefont {Chen}, \citenamefont {Lu}, \citenamefont {Shu}, \citenamefont
  {Peng}, \citenamefont {Wang},\ and\ \citenamefont {Pan}}]{Yin2017}%
  \BibitemOpen
  \bibfield  {author} {\bibinfo {author} {\bibfnamefont {J.}~\bibnamefont
  {Yin}}, \bibinfo {author} {\bibfnamefont {Y.}~\bibnamefont {Cao}}, \bibinfo
  {author} {\bibfnamefont {Y.-H.}\ \bibnamefont {Li}}, \bibinfo {author}
  {\bibfnamefont {S.-K.}\ \bibnamefont {Liao}}, \bibinfo {author}
  {\bibfnamefont {L.}~\bibnamefont {Zhang}}, \bibinfo {author} {\bibfnamefont
  {J.-G.}\ \bibnamefont {Ren}}, \bibinfo {author} {\bibfnamefont {W.-Q.}\
  \bibnamefont {Cai}}, \bibinfo {author} {\bibfnamefont {W.-Y.}\ \bibnamefont
  {Liu}}, \bibinfo {author} {\bibfnamefont {B.}~\bibnamefont {Li}}, \bibinfo
  {author} {\bibfnamefont {H.}~\bibnamefont {Dai}}, \bibinfo {author}
  {\bibfnamefont {G.-B.}\ \bibnamefont {Li}}, \bibinfo {author} {\bibfnamefont
  {Q.-M.}\ \bibnamefont {Lu}}, \bibinfo {author} {\bibfnamefont {Y.-H.}\
  \bibnamefont {Gong}}, \bibinfo {author} {\bibfnamefont {Y.}~\bibnamefont
  {Xu}}, \bibinfo {author} {\bibfnamefont {S.-L.}\ \bibnamefont {Li}}, \bibinfo
  {author} {\bibfnamefont {F.-Z.}\ \bibnamefont {Li}}, \bibinfo {author}
  {\bibfnamefont {Y.-Y.}\ \bibnamefont {Yin}}, \bibinfo {author} {\bibfnamefont
  {Z.-Q.}\ \bibnamefont {Jiang}}, \bibinfo {author} {\bibfnamefont
  {M.}~\bibnamefont {Li}}, \bibinfo {author} {\bibfnamefont {J.-J.}\
  \bibnamefont {Jia}}, \bibinfo {author} {\bibfnamefont {G.}~\bibnamefont
  {Ren}}, \bibinfo {author} {\bibfnamefont {D.}~\bibnamefont {He}}, \bibinfo
  {author} {\bibfnamefont {Y.-L.}\ \bibnamefont {Zhou}}, \bibinfo {author}
  {\bibfnamefont {X.-X.}\ \bibnamefont {Zhang}}, \bibinfo {author}
  {\bibfnamefont {N.}~\bibnamefont {Wang}}, \bibinfo {author} {\bibfnamefont
  {X.}~\bibnamefont {Chang}}, \bibinfo {author} {\bibfnamefont {Z.-C.}\
  \bibnamefont {Zhu}}, \bibinfo {author} {\bibfnamefont {N.-L.}\ \bibnamefont
  {Liu}}, \bibinfo {author} {\bibfnamefont {Y.-A.}\ \bibnamefont {Chen}},
  \bibinfo {author} {\bibfnamefont {C.-Y.}\ \bibnamefont {Lu}}, \bibinfo
  {author} {\bibfnamefont {R.}~\bibnamefont {Shu}}, \bibinfo {author}
  {\bibfnamefont {C.-Z.}\ \bibnamefont {Peng}}, \bibinfo {author}
  {\bibfnamefont {J.-Y.}\ \bibnamefont {Wang}}, \ and\ \bibinfo {author}
  {\bibfnamefont {J.-W.}\ \bibnamefont {Pan}},\ }\href {\doibase
  10.1126/science.aan3211} {\bibfield  {journal} {\bibinfo  {journal}
  {Science}\ }\textbf {\bibinfo {volume} {35}},\ \bibinfo {pages} {1140}
  (\bibinfo {year} {2017})}\BibitemShut {NoStop}%
\bibitem [{\citenamefont {Zurek}(2003)}]{Zurek2003}%
  \BibitemOpen
  \bibfield  {author} {\bibinfo {author} {\bibfnamefont {W.~H.}\ \bibnamefont
  {Zurek}},\ }\href {\doibase 10.1103/RevModPhys.75.715} {\bibfield  {journal}
  {\bibinfo  {journal} {Rev. Mod. Phys.}\ }\textbf {\bibinfo {volume} {75}},\
  \bibinfo {pages} {715} (\bibinfo {year} {2003})}\BibitemShut {NoStop}%
\bibitem [{\citenamefont {{Giovannetti}}\ \emph {et~al.}(2011)\citenamefont
  {{Giovannetti}}, \citenamefont {{Lloyd}},\ and\ \citenamefont
  {{Maccone}}}]{Giovannetti2011NaPho}%
  \BibitemOpen
  \bibfield  {author} {\bibinfo {author} {\bibfnamefont {V.}~\bibnamefont
  {{Giovannetti}}}, \bibinfo {author} {\bibfnamefont {S.}~\bibnamefont
  {{Lloyd}}}, \ and\ \bibinfo {author} {\bibfnamefont {L.}~\bibnamefont
  {{Maccone}}},\ }\href {\doibase 10.1038/nphoton.2011.35} {\bibfield
  {journal} {\bibinfo  {journal} {Nat. Photon.}\ }\textbf {\bibinfo {volume}
  {5}},\ \bibinfo {pages} {222} (\bibinfo {year} {2011})}\BibitemShut {NoStop}%
\bibitem [{\citenamefont {Marvian}\ and\ \citenamefont
  {Spekkens}(2016)}]{MarvianPhysRevA.94.052324}%
  \BibitemOpen
  \bibfield  {author} {\bibinfo {author} {\bibfnamefont {I.}~\bibnamefont
  {Marvian}}\ and\ \bibinfo {author} {\bibfnamefont {R.~W.}\ \bibnamefont
  {Spekkens}},\ }\href {\doibase 10.1103/PhysRevA.94.052324} {\bibfield
  {journal} {\bibinfo  {journal} {Phys. Rev. A}\ }\textbf {\bibinfo {volume}
  {94}},\ \bibinfo {pages} {052324} (\bibinfo {year} {2016})}\BibitemShut
  {NoStop}%
\bibitem [{\citenamefont {Hillery}(2016)}]{HilleryPhysRevA.93.012111}%
  \BibitemOpen
  \bibfield  {author} {\bibinfo {author} {\bibfnamefont {M.}~\bibnamefont
  {Hillery}},\ }\href {\doibase 10.1103/PhysRevA.93.012111} {\bibfield
  {journal} {\bibinfo  {journal} {Phys. Rev. A}\ }\textbf {\bibinfo {volume}
  {93}},\ \bibinfo {pages} {012111} (\bibinfo {year} {2016})}\BibitemShut
  {NoStop}%
\bibitem [{\citenamefont {Matera}\ \emph {et~al.}(2016)\citenamefont {Matera},
  \citenamefont {Egloff}, \citenamefont {Killoran},\ and\ \citenamefont
  {Plenio}}]{Matera2016}%
  \BibitemOpen
  \bibfield  {author} {\bibinfo {author} {\bibfnamefont {J.~M.}\ \bibnamefont
  {Matera}}, \bibinfo {author} {\bibfnamefont {D.}~\bibnamefont {Egloff}},
  \bibinfo {author} {\bibfnamefont {N.}~\bibnamefont {Killoran}}, \ and\
  \bibinfo {author} {\bibfnamefont {M.~B.}\ \bibnamefont {Plenio}},\ }\href
  {\doibase 10.1088/2058-9565/1/1/01LT01} {\bibfield  {journal} {\bibinfo
  {journal} {Quantum Sci. Technol.}\ }\textbf {\bibinfo {volume} {1}},\
  \bibinfo {pages} {01LT01} (\bibinfo {year} {2016})}\BibitemShut {NoStop}%
\bibitem [{\citenamefont {Huelga}\ and\ \citenamefont
  {Plenio}(2013)}]{Huelga2013}%
  \BibitemOpen
  \bibfield  {author} {\bibinfo {author} {\bibfnamefont {S.}~\bibnamefont
  {Huelga}}\ and\ \bibinfo {author} {\bibfnamefont {M.}~\bibnamefont
  {Plenio}},\ }\href {\doibase 10.1080/00405000.2013.829687} {\bibfield
  {journal} {\bibinfo  {journal} {Contemporary Physics}\ }\textbf {\bibinfo
  {volume} {54}},\ \bibinfo {pages} {181} (\bibinfo {year} {2013})}\BibitemShut
  {NoStop}%
\bibitem [{\citenamefont {Wootters}\ and\ \citenamefont
  {Zurek}(1982)}]{Wootters1982}%
  \BibitemOpen
  \bibfield  {author} {\bibinfo {author} {\bibfnamefont {W.~K.}\ \bibnamefont
  {Wootters}}\ and\ \bibinfo {author} {\bibfnamefont {W.~H.}\ \bibnamefont
  {Zurek}},\ }\href {\doibase 10.1038/299802a0} {\bibfield  {journal} {\bibinfo
   {journal} {Nature}\ }\textbf {\bibinfo {volume} {299}},\ \bibinfo {pages}
  {802} (\bibinfo {year} {1982})}\BibitemShut {NoStop}%
\bibitem [{\citenamefont {Streltsov}\ \emph {et~al.}(2015)\citenamefont
  {Streltsov}, \citenamefont {Singh}, \citenamefont {Dhar}, \citenamefont
  {Bera},\ and\ \citenamefont {Adesso}}]{Streltsov2015}%
  \BibitemOpen
  \bibfield  {author} {\bibinfo {author} {\bibfnamefont {A.}~\bibnamefont
  {Streltsov}}, \bibinfo {author} {\bibfnamefont {U.}~\bibnamefont {Singh}},
  \bibinfo {author} {\bibfnamefont {H.~S.}\ \bibnamefont {Dhar}}, \bibinfo
  {author} {\bibfnamefont {M.~N.}\ \bibnamefont {Bera}}, \ and\ \bibinfo
  {author} {\bibfnamefont {G.}~\bibnamefont {Adesso}},\ }\href {\doibase
  10.1103/PhysRevLett.115.020403} {\bibfield  {journal} {\bibinfo  {journal}
  {Phys. Rev. Lett.}\ }\textbf {\bibinfo {volume} {115}},\ \bibinfo {pages}
  {020403} (\bibinfo {year} {2015})}\BibitemShut {NoStop}%
\bibitem [{\citenamefont {Killoran}\ \emph {et~al.}(2016)\citenamefont
  {Killoran}, \citenamefont {Steinhoff},\ and\ \citenamefont
  {Plenio}}]{Killoran2016}%
  \BibitemOpen
  \bibfield  {author} {\bibinfo {author} {\bibfnamefont {N.}~\bibnamefont
  {Killoran}}, \bibinfo {author} {\bibfnamefont {F.~E.~S.}\ \bibnamefont
  {Steinhoff}}, \ and\ \bibinfo {author} {\bibfnamefont {M.~B.}\ \bibnamefont
  {Plenio}},\ }\href {\doibase 10.1103/PhysRevLett.116.080402} {\bibfield
  {journal} {\bibinfo  {journal} {Phys. Rev. Lett.}\ }\textbf {\bibinfo
  {volume} {116}},\ \bibinfo {pages} {080402} (\bibinfo {year}
  {2016})}\BibitemShut {NoStop}%
\bibitem [{\citenamefont {{Theurer}}\ \emph {et~al.}(2017)\citenamefont
  {{Theurer}}, \citenamefont {{Killoran}}, \citenamefont {{Egloff}},\ and\
  \citenamefont {{Plenio}}}]{Theurer2017}%
  \BibitemOpen
  \bibfield  {author} {\bibinfo {author} {\bibfnamefont {T.}~\bibnamefont
  {{Theurer}}}, \bibinfo {author} {\bibfnamefont {N.}~\bibnamefont
  {{Killoran}}}, \bibinfo {author} {\bibfnamefont {D.}~\bibnamefont
  {{Egloff}}}, \ and\ \bibinfo {author} {\bibfnamefont {M.~B.}\ \bibnamefont
  {{Plenio}}},\ }\href {https://arxiv.org/abs/1703.10943} {\bibfield  {journal}
  {\bibinfo  {journal} {arXiv:1703.10943}\ } (\bibinfo {year}
  {2017})}\BibitemShut {NoStop}%
\bibitem [{\citenamefont {Marvian}\ and\ \citenamefont
  {Spekkens}(2013)}]{Marvian2013}%
  \BibitemOpen
  \bibfield  {author} {\bibinfo {author} {\bibfnamefont {I.}~\bibnamefont
  {Marvian}}\ and\ \bibinfo {author} {\bibfnamefont {R.~W.}\ \bibnamefont
  {Spekkens}},\ }\href {\doibase 10.1088/1367-2630/15/3/033001} {\bibfield
  {journal} {\bibinfo  {journal} {New J. Phys.}\ }\textbf {\bibinfo {volume}
  {15}},\ \bibinfo {pages} {033001} (\bibinfo {year} {2013})}\BibitemShut
  {NoStop}%
\bibitem [{\citenamefont {{Chefles}}\ \emph {et~al.}(2004)\citenamefont
  {{Chefles}}, \citenamefont {{Jozsa}},\ and\ \citenamefont
  {{Winter}}}]{Chefles2004}%
  \BibitemOpen
  \bibfield  {author} {\bibinfo {author} {\bibfnamefont {A.}~\bibnamefont
  {{Chefles}}}, \bibinfo {author} {\bibfnamefont {R.}~\bibnamefont {{Jozsa}}},
  \ and\ \bibinfo {author} {\bibfnamefont {A.}~\bibnamefont {{Winter}}},\
  }\href {\doibase 10.1142/S0219749904000031} {\bibfield  {journal} {\bibinfo
  {journal} {Int. J. Quant. Inf.}\ }\textbf {\bibinfo {volume} {2}},\ \bibinfo
  {pages} {11} (\bibinfo {year} {2004})}\BibitemShut {NoStop}%
\bibitem [{\citenamefont {{Regula}}\ \emph {et~al.}(2017)\citenamefont
  {{Regula}}, \citenamefont {{Piani}}, \citenamefont {{Cianciaruso}},
  \citenamefont {{Bromley}}, \citenamefont {{Streltsov}},\ and\ \citenamefont
  {{Adesso}}}]{Regula1704.04153}%
  \BibitemOpen
  \bibfield  {author} {\bibinfo {author} {\bibfnamefont {B.}~\bibnamefont
  {{Regula}}}, \bibinfo {author} {\bibfnamefont {M.}~\bibnamefont {{Piani}}},
  \bibinfo {author} {\bibfnamefont {M.}~\bibnamefont {{Cianciaruso}}}, \bibinfo
  {author} {\bibfnamefont {T.~R.}\ \bibnamefont {{Bromley}}}, \bibinfo {author}
  {\bibfnamefont {A.}~\bibnamefont {{Streltsov}}}, \ and\ \bibinfo {author}
  {\bibfnamefont {G.}~\bibnamefont {{Adesso}}},\ }\href
  {https://arxiv.org/abs/1704.04153} {\bibfield  {journal} {\bibinfo  {journal}
  {arXiv:1704.04153}\ } (\bibinfo {year} {2017})}\BibitemShut {NoStop}%
\bibitem [{\citenamefont {Rana}\ \emph {et~al.}(2016)\citenamefont {Rana},
  \citenamefont {Parashar},\ and\ \citenamefont {Lewenstein}}]{Rana2016}%
  \BibitemOpen
  \bibfield  {author} {\bibinfo {author} {\bibfnamefont {S.}~\bibnamefont
  {Rana}}, \bibinfo {author} {\bibfnamefont {P.}~\bibnamefont {Parashar}}, \
  and\ \bibinfo {author} {\bibfnamefont {M.}~\bibnamefont {Lewenstein}},\
  }\href {\doibase 10.1103/PhysRevA.93.012110} {\bibfield  {journal} {\bibinfo
  {journal} {Phys. Rev. A}\ }\textbf {\bibinfo {volume} {93}},\ \bibinfo
  {pages} {012110} (\bibinfo {year} {2016})}\BibitemShut {NoStop}%
\bibitem [{\citenamefont {Chen}\ \emph {et~al.}(2016)\citenamefont {Chen},
  \citenamefont {Grogan}, \citenamefont {Johnston}, \citenamefont {Li},\ and\
  \citenamefont {Plosker}}]{Chen2016}%
  \BibitemOpen
  \bibfield  {author} {\bibinfo {author} {\bibfnamefont {J.}~\bibnamefont
  {Chen}}, \bibinfo {author} {\bibfnamefont {S.}~\bibnamefont {Grogan}},
  \bibinfo {author} {\bibfnamefont {N.}~\bibnamefont {Johnston}}, \bibinfo
  {author} {\bibfnamefont {C.-K.}\ \bibnamefont {Li}}, \ and\ \bibinfo {author}
  {\bibfnamefont {S.}~\bibnamefont {Plosker}},\ }\href {\doibase
  10.1103/PhysRevA.94.042313} {\bibfield  {journal} {\bibinfo  {journal} {Phys.
  Rev. A}\ }\textbf {\bibinfo {volume} {94}},\ \bibinfo {pages} {042313}
  (\bibinfo {year} {2016})}\BibitemShut {NoStop}%
\bibitem [{\citenamefont {Yu}\ \emph {et~al.}(2016)\citenamefont {Yu},
  \citenamefont {Zhang}, \citenamefont {Xu},\ and\ \citenamefont
  {Tong}}]{Yu2016}%
  \BibitemOpen
  \bibfield  {author} {\bibinfo {author} {\bibfnamefont {X.-D.}\ \bibnamefont
  {Yu}}, \bibinfo {author} {\bibfnamefont {D.-J.}\ \bibnamefont {Zhang}},
  \bibinfo {author} {\bibfnamefont {G.~F.}\ \bibnamefont {Xu}}, \ and\ \bibinfo
  {author} {\bibfnamefont {D.~M.}\ \bibnamefont {Tong}},\ }\href {\doibase
  10.1103/PhysRevA.94.060302} {\bibfield  {journal} {\bibinfo  {journal} {Phys.
  Rev. A}\ }\textbf {\bibinfo {volume} {94}},\ \bibinfo {pages} {060302}
  (\bibinfo {year} {2016})}\BibitemShut {NoStop}%
\bibitem [{\citenamefont {{Eisert}}(2001)}]{Eisert2006}%
  \BibitemOpen
  \bibfield  {author} {\bibinfo {author} {\bibfnamefont {J.}~\bibnamefont
  {{Eisert}}},\ }\emph {\bibinfo {title} {{Entanglement in quantum information
  theory}}},\ \href@noop {} {Ph.D. thesis},\ \bibinfo  {school} {University of
  Potsdam} (\bibinfo {year} {2001}),\ \Eprint
  {http://arxiv.org/abs/arXiv:quant-ph/0610253} {arXiv:quant-ph/0610253}
  \BibitemShut {NoStop}%
\bibitem [{\citenamefont {Wootters}(1998)}]{Wootters1998}%
  \BibitemOpen
  \bibfield  {author} {\bibinfo {author} {\bibfnamefont {W.~K.}\ \bibnamefont
  {Wootters}},\ }\href {\doibase 10.1103/PhysRevLett.80.2245} {\bibfield
  {journal} {\bibinfo  {journal} {Phys. Rev. Lett.}\ }\textbf {\bibinfo
  {volume} {80}},\ \bibinfo {pages} {2245} (\bibinfo {year}
  {1998})}\BibitemShut {NoStop}%
\bibitem [{\citenamefont {Shao}\ \emph {et~al.}(2015)\citenamefont {Shao},
  \citenamefont {Xi}, \citenamefont {Fan},\ and\ \citenamefont
  {Li}}]{Shao2015}%
  \BibitemOpen
  \bibfield  {author} {\bibinfo {author} {\bibfnamefont {L.-H.}\ \bibnamefont
  {Shao}}, \bibinfo {author} {\bibfnamefont {Z.}~\bibnamefont {Xi}}, \bibinfo
  {author} {\bibfnamefont {H.}~\bibnamefont {Fan}}, \ and\ \bibinfo {author}
  {\bibfnamefont {Y.}~\bibnamefont {Li}},\ }\href {\doibase
  10.1103/PhysRevA.91.042120} {\bibfield  {journal} {\bibinfo  {journal} {Phys.
  Rev. A}\ }\textbf {\bibinfo {volume} {91}},\ \bibinfo {pages} {042120}
  (\bibinfo {year} {2015})}\BibitemShut {NoStop}%
 \bibitem [{\citenamefont {Kiesel}\ \emph {et~al.}(2005)\citenamefont {Kiesel},
  \citenamefont {Schmid}, \citenamefont {Weber}, \citenamefont
  {Ursin},\ and\ \citenamefont
  {Weinfurter}}]{Kiesel2005}%
  \BibitemOpen
  \bibfield  {author} {\bibinfo {author} {\bibfnamefont {N.}\ \bibnamefont
  {Kiesel}}, \bibinfo {author} {\bibfnamefont {C.}~\bibnamefont {Schmid}},
  \bibinfo {author} {\bibfnamefont {U.}~\bibnamefont {Weber}}, \bibinfo
  {author} {\bibfnamefont {R.}~\bibnamefont {Ursin}}, \ and\ \bibinfo {author}
  {\bibfnamefont {H.}~\bibnamefont {Weinfurter}},\ }\href {\doibase
  10.1103/PhysRevLett.95.210505} {\bibfield  {journal} {\bibinfo  {journal}
  {Phys. Rev. Lett.}\ }\textbf {\bibinfo {volume} {95}},\ \bibinfo {pages}
  {210505} (\bibinfo {year} {2005})}\BibitemShut {NoStop}%
 \bibitem [{\citenamefont {Okamoto}\ \emph {et~al.}(2016)\citenamefont
  {Okamoto}, \citenamefont {Hofmann}, \citenamefont {Takeuchi},\ and\ \citenamefont
  {Sasaki}}]{Okamoto2005}%
  \BibitemOpen
  \bibfield  {author} {\bibinfo {author} {\bibfnamefont {R.}~\bibnamefont
  {Okamoto}}, \bibinfo {author} {\bibfnamefont {H.~F.}~\bibnamefont
  {Hofmann}}, \bibinfo {author} {\bibfnamefont {S.}~\bibnamefont {Takeuchi}},
   \ and\ \bibinfo {author}
  {\bibfnamefont {K.}~\bibnamefont {Sasaki}},\ }\href {\doibase
  10.1103/PhysRevLett.95.210506} {\bibfield  {journal} {\bibinfo  {journal}
  {Phys. Rev. Lett.}\ }\textbf {\bibinfo {volume} {95}},\ \bibinfo {pages}
  {210506} (\bibinfo {year} {2005})}\BibitemShut {NoStop}%
\bibitem [{\citenamefont {{Crespi}}\ \emph {et~al.}(2011)\citenamefont
  {{Crespi}}, \citenamefont {{Ramponi}}, \citenamefont {{Osellame}},
  \citenamefont {{Sansoni}}, \citenamefont {{Bongioanni}}, \citenamefont
  {{Sciarrino}}, \citenamefont {{Vallone}},\ and\ \citenamefont
  {{Mataloni}}}]{Crespi2011}%
  \BibitemOpen
  \bibfield  {author} {\bibinfo {author} {\bibfnamefont {A.}~\bibnamefont
  {{Crespi}}}, \bibinfo {author} {\bibfnamefont {R.}~\bibnamefont {{Ramponi}}},
  \bibinfo {author} {\bibfnamefont {R.}~\bibnamefont {{Osellame}}}, \bibinfo
  {author} {\bibfnamefont {L.}~\bibnamefont {{Sansoni}}}, \bibinfo {author}
  {\bibfnamefont {I.}~\bibnamefont {{Bongioanni}}}, \bibinfo {author}
  {\bibfnamefont {F.}~\bibnamefont {{Sciarrino}}}, \bibinfo {author}
  {\bibfnamefont {G.}~\bibnamefont {{Vallone}}}, \ and\ \bibinfo {author}
  {\bibfnamefont {P.}~\bibnamefont {{Mataloni}}},\ }\href {\doibase
  10.1038/ncomms1570} {\bibfield  {journal} {\bibinfo  {journal} {Nature
  Commun.}\ }\textbf {\bibinfo {volume} {2}},\ \bibinfo {eid} {566} (\bibinfo
  {year} {2011})}\BibitemShut {NoStop}%
\bibitem [{\citenamefont {Kim}(2003)}]{Kim2003}%
  \BibitemOpen
  \bibfield  {author} {\bibinfo {author} {\bibfnamefont {Y.-H.}\ \bibnamefont
  {Kim}},\ }\href {\doibase 10.1103/PhysRevA.68.013804} {\bibfield  {journal}
  {\bibinfo  {journal} {Phys. Rev. A}\ }\textbf {\bibinfo {volume} {68}},\
  \bibinfo {pages} {013804} (\bibinfo {year} {2003})}\BibitemShut {NoStop}%
\bibitem [{\citenamefont {Kwiat}\ \emph {et~al.}(1995)\citenamefont {Kwiat},
  \citenamefont {Mattle}, \citenamefont {Weinfurter}, \citenamefont
  {Zeilinger}, \citenamefont {Sergienko},\ and\ \citenamefont
  {Shih}}]{Kwiat1995}%
  \BibitemOpen
  \bibfield  {author} {\bibinfo {author} {\bibfnamefont {P.~G.}\ \bibnamefont
  {Kwiat}}, \bibinfo {author} {\bibfnamefont {K.}~\bibnamefont {Mattle}},
  \bibinfo {author} {\bibfnamefont {H.}~\bibnamefont {Weinfurter}}, \bibinfo
  {author} {\bibfnamefont {A.}~\bibnamefont {Zeilinger}}, \bibinfo {author}
  {\bibfnamefont {A.~V.}\ \bibnamefont {Sergienko}}, \ and\ \bibinfo {author}
  {\bibfnamefont {Y.}~\bibnamefont {Shih}},\ }\href {\doibase
  10.1103/PhysRevLett.75.4337} {\bibfield  {journal} {\bibinfo  {journal}
  {Phys. Rev. Lett.}\ }\textbf {\bibinfo {volume} {75}},\ \bibinfo {pages}
  {4337} (\bibinfo {year} {1995})}\BibitemShut {NoStop}%
\bibitem [{\citenamefont {Hong}\ \emph {et~al.}(1987)\citenamefont {Hong},
  \citenamefont {Ou},\ and\ \citenamefont {Mandel}}]{Hong1987}%
  \BibitemOpen
  \bibfield  {author} {\bibinfo {author} {\bibfnamefont {C.~K.}\ \bibnamefont
  {Hong}}, \bibinfo {author} {\bibfnamefont {Z.~Y.}\ \bibnamefont {Ou}}, \ and\
  \bibinfo {author} {\bibfnamefont {L.}~\bibnamefont {Mandel}},\ }\href
  {\doibase 10.1103/PhysRevLett.59.2044} {\bibfield  {journal} {\bibinfo
  {journal} {Phys. Rev. Lett.}\ }\textbf {\bibinfo {volume} {59}},\ \bibinfo
  {pages} {2044} (\bibinfo {year} {1987})}\BibitemShut {NoStop}%
\bibitem [{\citenamefont {Hofmann}(2005)}]{Hofmann2005}%
  \BibitemOpen
  \bibfield  {author} {\bibinfo {author} {\bibfnamefont {H.~F.}\ \bibnamefont
  {Hofmann}},\ }\href {\doibase 10.1103/PhysRevLett.94.160504} {\bibfield
  {journal} {\bibinfo  {journal} {Phys. Rev. Lett.}\ }\textbf {\bibinfo
  {volume} {94}},\ \bibinfo {pages} {160504} (\bibinfo {year}
  {2005})}\BibitemShut {NoStop}%
\bibitem [{\citenamefont {James}\ \emph {et~al.}(2001)\citenamefont {James},
  \citenamefont {Kwiat}, \citenamefont {Munro},\ and\ \citenamefont
  {White}}]{James2001}%
  \BibitemOpen
  \bibfield  {author} {\bibinfo {author} {\bibfnamefont {D.~F.~V.}\
  \bibnamefont {James}}, \bibinfo {author} {\bibfnamefont {P.~G.}\ \bibnamefont
  {Kwiat}}, \bibinfo {author} {\bibfnamefont {W.~J.}\ \bibnamefont {Munro}}, \
  and\ \bibinfo {author} {\bibfnamefont {A.~G.}\ \bibnamefont {White}},\ }\href
  {\doibase 10.1103/PhysRevA.64.052312} {\bibfield  {journal} {\bibinfo
  {journal} {Phys. Rev. A}\ }\textbf {\bibinfo {volume} {64}},\ \bibinfo
  {pages} {052312} (\bibinfo {year} {2001})}\BibitemShut {NoStop}%
\bibitem [{\citenamefont {{Wu}}\ \emph {et~al.}(2017)\citenamefont {{Wu}},
  \citenamefont {{Hou}}, \citenamefont {{Zhao}}, \citenamefont {{Xiang}},
  \citenamefont {{Li}}, \citenamefont {{Guo}}, \citenamefont {{Ma}},
  \citenamefont {{He}}, \citenamefont {{Thompson}},\ and\ \citenamefont
  {{Gu}}}]{Wu1710.01738}%
  \BibitemOpen
  \bibfield  {author} {\bibinfo {author} {\bibfnamefont {K.-D.}\ \bibnamefont
  {{Wu}}}, \bibinfo {author} {\bibfnamefont {Z.}~\bibnamefont {{Hou}}},
  \bibinfo {author} {\bibfnamefont {Y.-Y.}\ \bibnamefont {{Zhao}}}, \bibinfo
  {author} {\bibfnamefont {G.-Y.}\ \bibnamefont {{Xiang}}}, \bibinfo {author}
  {\bibfnamefont {C.-F.}\ \bibnamefont {{Li}}}, \bibinfo {author}
  {\bibfnamefont {G.-C.}\ \bibnamefont {{Guo}}}, \bibinfo {author}
  {\bibfnamefont {J.}~\bibnamefont {{Ma}}}, \bibinfo {author} {\bibfnamefont
  {Q.-Y.}\ \bibnamefont {{He}}}, \bibinfo {author} {\bibfnamefont
  {J.}~\bibnamefont {{Thompson}}}, \ and\ \bibinfo {author} {\bibfnamefont
  {M.}~\bibnamefont {{Gu}}},\ }\href {https://arxiv.org/abs/1710.01738}
  {\bibfield  {journal} {\bibinfo  {journal} {arXiv:1710.01738}\ } (\bibinfo
  {year} {2017})}\BibitemShut {NoStop}%
\bibitem [{\citenamefont {Chitambar}\ \emph {et~al.}(2016)\citenamefont
  {Chitambar}, \citenamefont {Streltsov}, \citenamefont {Rana}, \citenamefont
  {Bera}, \citenamefont {Adesso},\ and\ \citenamefont
  {Lewenstein}}]{Chitambar2016}%
  \BibitemOpen
  \bibfield  {author} {\bibinfo {author} {\bibfnamefont {E.}~\bibnamefont
  {Chitambar}}, \bibinfo {author} {\bibfnamefont {A.}~\bibnamefont
  {Streltsov}}, \bibinfo {author} {\bibfnamefont {S.}~\bibnamefont {Rana}},
  \bibinfo {author} {\bibfnamefont {M.~N.}\ \bibnamefont {Bera}}, \bibinfo
  {author} {\bibfnamefont {G.}~\bibnamefont {Adesso}}, \ and\ \bibinfo {author}
  {\bibfnamefont {M.}~\bibnamefont {Lewenstein}},\ }\href {\doibase
  10.1103/PhysRevLett.116.070402} {\bibfield  {journal} {\bibinfo  {journal}
  {Phys. Rev. Lett.}\ }\textbf {\bibinfo {volume} {116}},\ \bibinfo {pages}
  {070402} (\bibinfo {year} {2016})}\BibitemShut {NoStop}%
\bibitem [{\citenamefont {Streltsov}\ \emph {et~al.}(2017)\citenamefont
  {Streltsov}, \citenamefont {Rana}, \citenamefont {Bera},\ and\ \citenamefont
  {Lewenstein}}]{Streltsov2017}%
  \BibitemOpen
  \bibfield  {author} {\bibinfo {author} {\bibfnamefont {A.}~\bibnamefont
  {Streltsov}}, \bibinfo {author} {\bibfnamefont {S.}~\bibnamefont {Rana}},
  \bibinfo {author} {\bibfnamefont {M.~N.}\ \bibnamefont {Bera}}, \ and\
  \bibinfo {author} {\bibfnamefont {M.}~\bibnamefont {Lewenstein}},\ }\href
  {\doibase 10.1103/PhysRevX.7.011024} {\bibfield  {journal} {\bibinfo
  {journal} {Phys. Rev. X}\ }\textbf {\bibinfo {volume} {7}},\ \bibinfo {pages}
  {011024} (\bibinfo {year} {2017})}\BibitemShut {NoStop}%
\bibitem [{\citenamefont {Modi}\ \emph {et~al.}(2012)\citenamefont {Modi},
  \citenamefont {Brodutch}, \citenamefont {Cable}, \citenamefont {Paterek},\
  and\ \citenamefont {Vedral}}]{Modi2012}%
  \BibitemOpen
  \bibfield  {author} {\bibinfo {author} {\bibfnamefont {K.}~\bibnamefont
  {Modi}}, \bibinfo {author} {\bibfnamefont {A.}~\bibnamefont {Brodutch}},
  \bibinfo {author} {\bibfnamefont {H.}~\bibnamefont {Cable}}, \bibinfo
  {author} {\bibfnamefont {T.}~\bibnamefont {Paterek}}, \ and\ \bibinfo
  {author} {\bibfnamefont {V.}~\bibnamefont {Vedral}},\ }\href {\doibase
  10.1103/RevModPhys.84.1655} {\bibfield  {journal} {\bibinfo  {journal} {Rev.
  Mod. Phys.}\ }\textbf {\bibinfo {volume} {84}},\ \bibinfo {pages} {1655}
  (\bibinfo {year} {2012})}\BibitemShut {NoStop}%
\bibitem [{\citenamefont {Ma}\ \emph {et~al.}(2016)\citenamefont {Ma},
  \citenamefont {Yadin}, \citenamefont {Girolami}, \citenamefont {Vedral},\
  and\ \citenamefont {Gu}}]{Ma2016}%
  \BibitemOpen
  \bibfield  {author} {\bibinfo {author} {\bibfnamefont {J.}~\bibnamefont
  {Ma}}, \bibinfo {author} {\bibfnamefont {B.}~\bibnamefont {Yadin}}, \bibinfo
  {author} {\bibfnamefont {D.}~\bibnamefont {Girolami}}, \bibinfo {author}
  {\bibfnamefont {V.}~\bibnamefont {Vedral}}, \ and\ \bibinfo {author}
  {\bibfnamefont {M.}~\bibnamefont {Gu}},\ }\href {\doibase
  10.1103/PhysRevLett.116.160407} {\bibfield  {journal} {\bibinfo  {journal}
  {Phys. Rev. Lett.}\ }\textbf {\bibinfo {volume} {116}},\ \bibinfo {pages}
  {160407} (\bibinfo {year} {2016})}\BibitemShut {NoStop}%
\bibitem [{\citenamefont {Winter}\ and\ \citenamefont
  {Yang}(2016)}]{Winter2016}%
  \BibitemOpen
  \bibfield  {author} {\bibinfo {author} {\bibfnamefont {A.}~\bibnamefont
  {Winter}}\ and\ \bibinfo {author} {\bibfnamefont {D.}~\bibnamefont {Yang}},\
  }\href {\doibase 10.1103/PhysRevLett.116.120404} {\bibfield  {journal}
  {\bibinfo  {journal} {Phys. Rev. Lett.}\ }\textbf {\bibinfo {volume} {116}},\
  \bibinfo {pages} {120404} (\bibinfo {year} {2016})}\BibitemShut {NoStop}%
\bibitem [{\citenamefont {Yadin}\ \emph {et~al.}(2016)\citenamefont {Yadin},
  \citenamefont {Ma}, \citenamefont {Girolami}, \citenamefont {Gu},\ and\
  \citenamefont {Vedral}}]{Yadin2016}%
  \BibitemOpen
  \bibfield  {author} {\bibinfo {author} {\bibfnamefont {B.}~\bibnamefont
  {Yadin}}, \bibinfo {author} {\bibfnamefont {J.}~\bibnamefont {Ma}}, \bibinfo
  {author} {\bibfnamefont {D.}~\bibnamefont {Girolami}}, \bibinfo {author}
  {\bibfnamefont {M.}~\bibnamefont {Gu}}, \ and\ \bibinfo {author}
  {\bibfnamefont {V.}~\bibnamefont {Vedral}},\ }\href {\doibase
  10.1103/PhysRevX.6.041028} {\bibfield  {journal} {\bibinfo  {journal} {Phys.
  Rev. X}\ }\textbf {\bibinfo {volume} {6}},\ \bibinfo {pages} {041028}
  (\bibinfo {year} {2016})}\BibitemShut {NoStop}%
\bibitem [{\citenamefont {Wei}\ and\ \citenamefont {Goldbart}(2003)}]{Wei2003}%
  \BibitemOpen
  \bibfield  {author} {\bibinfo {author} {\bibfnamefont {T.-C.}\ \bibnamefont
  {Wei}}\ and\ \bibinfo {author} {\bibfnamefont {P.~M.}\ \bibnamefont
  {Goldbart}},\ }\href {\doibase 10.1103/PhysRevA.68.042307} {\bibfield
  {journal} {\bibinfo  {journal} {Phys. Rev. A}\ }\textbf {\bibinfo {volume}
  {68}},\ \bibinfo {pages} {042307} (\bibinfo {year} {2003})}\BibitemShut
  {NoStop}%
\bibitem [{\citenamefont {Streltsov}\ \emph {et~al.}(2010)\citenamefont
  {Streltsov}, \citenamefont {Kampermann},\ and\ \citenamefont
  {Bru\ss}}]{Streltsov2010}%
  \BibitemOpen
  \bibfield  {author} {\bibinfo {author} {\bibfnamefont {A.}~\bibnamefont
  {Streltsov}}, \bibinfo {author} {\bibfnamefont {H.}~\bibnamefont
  {Kampermann}}, \ and\ \bibinfo {author} {\bibfnamefont {D.}~\bibnamefont
  {Bru\ss}},\ }\href {\doibase 10.1088/1367-2630/12/12/123004} {\bibfield
  {journal} {\bibinfo  {journal} {New J. Phys.}\ }\textbf {\bibinfo {volume}
  {12}},\ \bibinfo {pages} {123004} (\bibinfo {year} {2010})}\BibitemShut
  {NoStop}%
\end{thebibliography}
%

\appendix

\section{\label{sec:Superposition-free-unitaries} Proof of Theorem~\ref{thm:2}}

In the following, we will characterize all superposition-free unitaries
acting on two qubits. In particular, we will show that any superposition-free
unitary in this framework can be decomposed into a sequence of elementary
unitaries $V$ and $W$, given in Eqs.~(\ref{eq:V}) and (\ref{eq:W})
of the main text. An important ingredient for our
proof is the following lemma \cite{Chefles2004,Marvian2013,Killoran2016}. 
\begin{lem}
\label{lem:1}For two sets of states $\left\{ \ket{\psi_{i}}\right\} _{i=1}^{N}$
and $\left\{ \ket{\phi_{i}}\right\} _{i=1}^{N}$ there exists a unitary
operation such that $U\ket{\psi_{i}}=\ket{\phi_{i}}$ for all $i$
if and only if $\braket{\psi_{i}|\psi_{j}}=\braket{\phi_{i}|\phi_{j}}$
holds true for all $i$ and $j$. 
\end{lem}
In general, a superposition-free unitary $U$ acts on a superposition-free
state $\ket{c_{k}}\ket{c_{l}}$ as follows: 
\begin{equation}
U\ket{c_{k}}\ket{c_{l}}=e^{i\phi_{kl}}\ket{c_{m}}\ket{c_{n}},
\end{equation}
where the possible final states $e^{i\phi_{kl}}\ket{c_{m}}\ket{c_{n}}$
are constrained by Lemma~\ref{lem:1}. As we will see in the following,
there exist 8 classes of superposition-free unitaries. For each of
those classes we will find a decomposition into the elementary operations
$V$ and $W$. 

\textbf{Class 1.} We start with the most simple transformation, corresponding
to the situation where an initial superposition-free state remains
unchanged (up to a possible phase): \begin{subequations} 
\begin{align}
\ket{c_{0}}\ket{c_{0}} & \rightarrow e^{i\phi_{00}}\ket{c_{0}}\ket{c_{0}},\\
\ket{c_{1}}\ket{c_{1}} & \rightarrow e^{i\phi_{11}}\ket{c_{1}}\ket{c_{1}},\\
\ket{c_{0}}\ket{c_{1}} & \rightarrow e^{i\phi_{01}}\ket{c_{0}}\ket{c_{1}},\\
\ket{c_{1}}\ket{c_{0}} & \rightarrow e^{i\phi_{10}}\ket{c_{1}}\ket{c_{0}}.
\end{align}
\end{subequations} Note that by Lemma~\ref{lem:1}, all phases $e^{i\phi_{kl}}$
must be equal. It is straightforward to see that this transformation
corresponds to $V^{2}$.

\textbf{Class 2.} We now consider the transformation \begin{subequations}
\begin{align}
\ket{c_{0}}\ket{c_{0}} & \rightarrow e^{i\phi_{00}}\ket{c_{0}}\ket{c_{0}},\\
\ket{c_{1}}\ket{c_{1}} & \rightarrow e^{i\phi_{11}}\ket{c_{1}}\ket{c_{1}},\\
\ket{c_{0}}\ket{c_{1}} & \rightarrow e^{i\phi_{01}}\ket{c_{1}}\ket{c_{0}},\\
\ket{c_{1}}\ket{c_{0}} & \rightarrow e^{i\phi_{10}}\ket{c_{0}}\ket{c_{1}}.
\end{align}
\end{subequations} By applying Lemma~\ref{lem:1}, we see that --
similar as in the previous case -- all phases $e^{i\phi_{kl}}$ must
be equal. This transformation corresponds to the swap unitary $V$.

\textbf{Class 3.} The next transformation that we will consider has
the following form: \begin{subequations} 
\begin{align}
\ket{c_{0}}\ket{c_{0}} & \rightarrow e^{i\phi_{00}}\ket{c_{1}}\ket{c_{1}},\\
\ket{c_{1}}\ket{c_{1}} & \rightarrow e^{i\phi_{11}}\ket{c_{0}}\ket{c_{0}},\\
\ket{c_{0}}\ket{c_{1}} & \rightarrow e^{i\phi_{01}}\ket{c_{0}}\ket{c_{1}},\\
\ket{c_{1}}\ket{c_{0}} & \rightarrow e^{i\phi_{10}}\ket{c_{1}}\ket{c_{0}}.
\end{align}
\end{subequations} Up to an overall phase, the phases $e^{i\phi_{kl}}$
are fixed by Lemma~\ref{lem:1} as follows: \begin{subequations}
\label{eq:phases-3} 
\begin{align}
e^{i\phi_{00}} & =1,\\
e^{i\phi_{01}} & =e^{i\phi_{10}}=e^{i\frac{\phi_{11}}{2}}=\frac{\braket{c_{0}|c_{1}}}{\braket{c_{1}|c_{0}}}.
\end{align}
\end{subequations} This transformation corresponds to the unitary
$WVW$.

\textbf{Class 4.} In the next step we consider the following transformation:
\begin{subequations} 
\begin{align}
\ket{c_{0}}\ket{c_{0}} & \rightarrow e^{i\phi_{00}}\ket{c_{1}}\ket{c_{1}},\\
\ket{c_{1}}\ket{c_{1}} & \rightarrow e^{i\phi_{11}}\ket{c_{0}}\ket{c_{0}},\\
\ket{c_{0}}\ket{c_{1}} & \rightarrow e^{i\phi_{01}}\ket{c_{1}}\ket{c_{0}},\\
\ket{c_{1}}\ket{c_{0}} & \rightarrow e^{i\phi_{10}}\ket{c_{0}}\ket{c_{1}}.
\end{align}
\end{subequations} It can be verified by inspection that (up to an
overall phase), Lemma~\ref{lem:1} fixes the phases $e^{i\phi_{kl}}$
in the same way as in Eq.~(\ref{eq:phases-3}). Note that this transformation
corresponds to the transformation of Class 3, followed by a swap.
Thus, it corresponds to the unitary $(VW)^{2}$.

\textbf{Class 5.} We now consider the transformation \begin{subequations}
\begin{align}
\ket{c_{0}}\ket{c_{0}} & \rightarrow e^{i\phi_{00}}\ket{c_{1}}\ket{c_{0}},\\
\ket{c_{1}}\ket{c_{1}} & \rightarrow e^{i\phi_{11}}\ket{c_{0}}\ket{c_{1}},\\
\ket{c_{0}}\ket{c_{1}} & \rightarrow e^{i\phi_{01}}\ket{c_{1}}\ket{c_{1}},\\
\ket{c_{1}}\ket{c_{0}} & \rightarrow e^{i\phi_{10}}\ket{c_{0}}\ket{c_{0}}.
\end{align}
\end{subequations} Up to an overall phase, Lemma~\ref{lem:1} fixes
the phases $e^{i\phi_{kl}}$ as follows: \begin{subequations} \label{eq:phases-5}
\begin{align}
e^{i\phi_{00}} & =e^{i\phi_{01}}=1,\\
e^{i\phi_{11}} & =e^{i\phi_{10}}=\frac{\braket{c_{0}|c_{1}}}{\braket{c_{1}|c_{0}}}.
\end{align}
\end{subequations} This transformation corresponds to the unitary
$W$.

\textbf{Class 6.} In the next step we consider the transformation
\begin{subequations} 
\begin{align}
\ket{c_{0}}\ket{c_{0}} & \rightarrow e^{i\phi_{00}}\ket{c_{1}}\ket{c_{0}},\\
\ket{c_{1}}\ket{c_{1}} & \rightarrow e^{i\phi_{11}}\ket{c_{0}}\ket{c_{1}},\\
\ket{c_{0}}\ket{c_{1}} & \rightarrow e^{i\phi_{01}}\ket{c_{0}}\ket{c_{0}},\\
\ket{c_{1}}\ket{c_{0}} & \rightarrow e^{i\phi_{10}}\ket{c_{1}}\ket{c_{1}}.
\end{align}
\end{subequations} By applying Lemma~\ref{lem:1}, we see that the
phases $e^{i\phi_{kl}}$ are fixed as follows: \begin{subequations}
\label{eq:phases-6} 
\begin{align}
e^{i\phi_{00}} & =e^{i\phi_{10}}=1,\\
e^{i\phi_{11}} & =e^{i\phi_{01}}=\frac{\braket{c_{0}|c_{1}}}{\braket{c_{1}|c_{0}}}.
\end{align}
\end{subequations} As can be verified by inspection, this transformation
corresponds to the unitary $WV$.

\textbf{Class 7.} The next transformation that we will consider has
the following form: \begin{subequations} 
\begin{align}
\ket{c_{0}}\ket{c_{0}} & \rightarrow e^{i\phi_{00}}\ket{c_{0}}\ket{c_{1}},\\
\ket{c_{1}}\ket{c_{1}} & \rightarrow e^{i\phi_{11}}\ket{c_{1}}\ket{c_{0}},\\
\ket{c_{0}}\ket{c_{1}} & \rightarrow e^{i\phi_{01}}\ket{c_{0}}\ket{c_{0}},\\
\ket{c_{1}}\ket{c_{0}} & \rightarrow e^{i\phi_{10}}\ket{c_{1}}\ket{c_{1}}.
\end{align}
\end{subequations} Up to an overall phase, Lemma~\ref{lem:1} fixes
the phases $e^{i\phi_{kl}}$ as in Eqs.~(\ref{eq:phases-6}). This
transformation corresponds to the transformation of Class 6 followed
by a swap, and the corresponding unitary is $VWV$.

\textbf{Class 8.} Our final transformation has the following form:
\begin{subequations} 
\begin{align}
\ket{c_{0}}\ket{c_{0}} & \rightarrow e^{i\phi_{00}}\ket{c_{0}}\ket{c_{1}},\\
\ket{c_{1}}\ket{c_{1}} & \rightarrow e^{i\phi_{11}}\ket{c_{1}}\ket{c_{0}},\\
\ket{c_{0}}\ket{c_{1}} & \rightarrow e^{i\phi_{01}}\ket{c_{1}}\ket{c_{1}},\\
\ket{c_{1}}\ket{c_{0}} & \rightarrow e^{i\phi_{10}}\ket{c_{0}}\ket{c_{0}}.
\end{align}
\end{subequations} Up to an overall phase, Lemma~\ref{lem:1} fixes
the phases $e^{i\phi_{kl}}$ as in Eq.~(\ref{eq:phases-5}). This
transformation corresponds to the transformation of Class 5 followed
by a swap, and the corresponding unitary is $VW$.

As we will discuss in the following, these eight classes indeed characterize
all superposition-free unitaries on two qubits. This can be seen by
inspection, applying Lemma~\ref{lem:1} to all the remaining permutations
of the superposition-free states. As an example, consider the following
transition: \begin{subequations} \label{eq:CNOT} 
\begin{align}
\ket{c_{0}}\ket{c_{0}} & \rightarrow e^{i\phi_{00}}\ket{c_{0}}\ket{c_{0}},\label{eq:violation-1}\\
\ket{c_{1}}\ket{c_{1}} & \rightarrow e^{i\phi_{11}}\ket{c_{1}}\ket{c_{0}},\label{eq:violation-2}\\
\ket{c_{0}}\ket{c_{1}} & \rightarrow e^{i\phi_{01}}\ket{c_{0}}\ket{c_{1}},\\
\ket{c_{1}}\ket{c_{0}} & \rightarrow e^{i\phi_{10}}\ket{c_{1}}\ket{c_{1}}.
\end{align}
\end{subequations} Transition of this form can be regarded as CNOT
operation in the resource theory of superposition, as (up to a phase)
the state of the second qubit is flipped between $\ket{c_{0}}$ and
$\ket{c_{1}}$, conditioned on the first qubit being in one of these
states.

The transition in Eqs.~(\ref{eq:CNOT}) is not covered by the above
classes, and it is indeed impossible via unitary operations. If such
a transition was possible via unitaries, this would lead to a violation
of Lemma~\ref{lem:1}. In particular, Lemma~\ref{lem:1} together
with Eqs.~(\ref{eq:violation-1}) and (\ref{eq:violation-2}) implies
that 
\begin{equation}
\braket{c_{0}|c_{1}}^{2}=e^{i(\phi_{11}-\phi_{00})}\braket{c_{0}|c_{1}},
\end{equation}
which cannot be true for any choice of the phases $e^{i\phi_{00}}$
and $e^{i\phi_{11}}$ in the considered range $0<|\!\braket{c_{0}|c_{1}}\!|<1$.
By similar arguments, all transitions which are not covered by the
above classes can be ruled out, and the proof is complete.

\section{\label{sec:trace-norm}Proof of Theorem \ref{thm:trace-norm}}

In the following, we will use results from \cite{Chen2016}, where
the authors provided an important link between $E_{\mathrm{t}}$ and
$C_{\mathrm{t}}$. In particular, theorems 2 and 3 in \cite{Chen2016}
imply the following equality: 
\begin{equation}
E_{\mathrm{t}}\left(\frac{1}{d}\sum_{i,j=0}^{d-1}\ket{ii}\!\bra{jj}\right)=C_{\mathrm{t}}\left(\frac{1}{d}\sum_{i,j=0}^{d-1}\ket{i}\!\bra{j}\right)=2-\frac{2}{d}.\label{eq:EtCt}
\end{equation}
Equipped with these tools we are now in position to prove Theorem~\ref{thm:trace-norm}
of the main text.

We will consider the bipartite state 
\begin{equation}
\rho=\frac{p}{2}\sum_{i,j=0}^{1}\ket{ii}\!\bra{jj}+\frac{1-p}{3}\sum_{k,l=2}^{4}\ket{kk}\!\bra{ll}\label{eq:violation-state}
\end{equation}
with probability $0\leq p\leq1$. Consider now local measurement on
the first party with Kraus operators 
\begin{equation}
K_{1}=\sum_{i=0}^{1}\ket{i}\!\bra{i}\otimes\openone,\,\,\,\,\,\,\,K_{2}=\sum_{j=2}^{4}\ket{j}\!\bra{j}\otimes\openone.
\end{equation}
It is straightforward to check that the corresponding measurement
probabilities take the form 
\begin{align}
q_{1} & =\mathrm{Tr}\left[K_{1}\rho K_{1}^{\dagger}\right]=p,\\
q_{2} & =\mathrm{Tr}\left[K_{2}\rho K_{2}^{\dagger}\right]=1-p.
\end{align}
Moreover, the post-measurement states are given as 
\begin{align}
\sigma_{1} & =\frac{K_{1}\rho K_{1}^{\dagger}}{p_{1}}=\frac{1}{2}\sum_{i,j=0}^{1}\ket{ii}\!\bra{jj},\\
\sigma_{2} & =\frac{K_{2}\rho K_{2}^{\dagger}}{p_{2}}=\frac{1}{3}\sum_{k,l=2}^{4}\ket{kk}\!\bra{ll}.
\end{align}

\begin{figure}
\includegraphics[width=1\columnwidth]{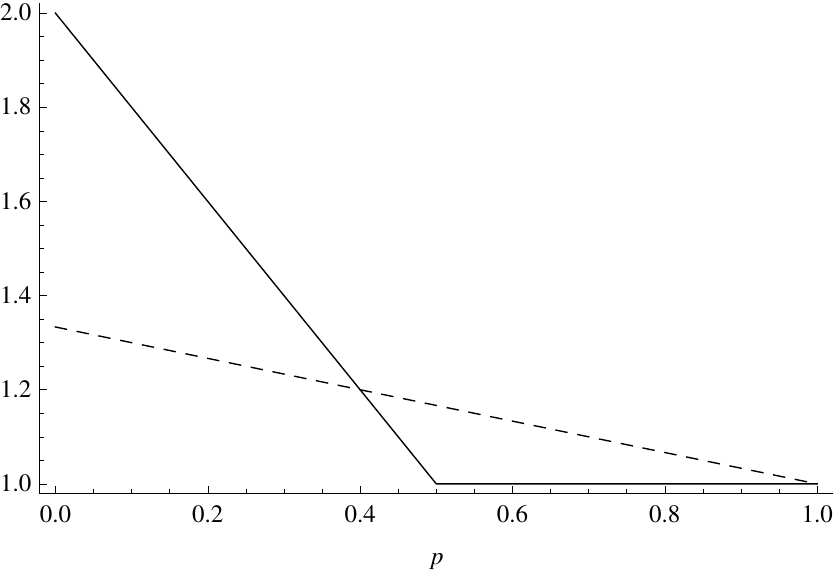}
\caption{\label{fig:violation}Violation of strong monotonicity of trace norm
entanglement for the state $\rho$ given in Eq.~(\ref{eq:violation-state}).
Solid line shows an upper bound on the trace norm entanglement of
$\rho$. Dashed line shows the average entanglement $q_{1}E_{\mathrm{t}}(\sigma_{1})+q_{2}E_{\mathrm{t}}(\sigma_{2})$
after a suitable local measurement. Violation of strong monotonicity
is obtained in the range $0.4<p<1$.}
\end{figure}

We will now complete the proof of the theorem by showing that for
a suitable choice of the probability $p$ it holds that 
\begin{equation}
q_{1}E_{\mathrm{t}}\left(\sigma_{1}\right)+q_{2}E_{\mathrm{t}}\left(\sigma_{2}\right)>E_{\mathrm{t}}\left(\rho\right).\label{eq:violation}
\end{equation}
For this, we define the separable state $\delta=\frac{1}{2}\sum_{i=0}^{1}\ket{ii}\!\bra{ii}$,
and note that it provides an upper bound on the trace norm entanglement,
i.e., $E_{\mathrm{t}}(\rho)\leq||\rho-\delta||_{1}$. Moreover, it
is straightforward to verify that 
\begin{equation}
||\rho-\delta||_{1}=\begin{cases}
2-2p & \mathrm{for\,}p<\frac{1}{2},\\
1 & \mathrm{for\,}p\geq\frac{1}{2}.
\end{cases}
\end{equation}
On the other hand, using Eq.~(\ref{eq:EtCt}) we obtain 
\begin{align}
E_{\mathrm{t}}(\sigma_{1}) & =1,\,\,\,\,\,\,\,E_{\mathrm{t}}(\sigma_{2})=\frac{4}{3}.
\end{align}
Using these results, we immediately see that Eq.~(\ref{eq:violation})
is fulfilled for $0.4<p<1$, see also Fig.~\ref{fig:violation}.

\section{\label{sec:concurrence}Activation of $\ell_{1}$-norm coherence
\protect \\
into concurrence}

We will now show that the inequality 
\begin{equation}
E(\Lambda_{i}[\rho\otimes\sigma_{i}])\leq C(\rho)\label{eq:EntCoh-2}
\end{equation}
holds for $\ell_{1}$-norm coherence $C$ and concurrence $E$, where
$\rho$ and $\sigma_{i}$ are single-qubit states, and $\Lambda_{i}$
is a bipartite incoherent operation. Moreover, we will also see that
equality in Eq.~(\ref{eq:EntCoh-2}) is achieved if $\Lambda_{i}$
is a CNOT gate. 

For proving the statement, we first recall the definition of geometric
entanglement \cite{Wei2003,Streltsov2010} and geometric coherence
\cite{Streltsov2015}
\begin{align}
E_{\mathrm{g}}(\rho) & =1-\max_{\sigma\in\mathcal{S}}F(\rho,\sigma),\\
C_{\mathrm{g}}(\rho) & =1-\max_{\sigma\in\mathcal{I}}F(\rho,\sigma)
\end{align}
with fidelity $F(\rho,\sigma)=||\sqrt{\rho}\sqrt{\sigma}||_{1}^{2}$.
Note that these quantities fulfill Eq.~(\ref{eq:EntCoh-2}), and
equality is attained if $\Lambda_{i}$ is a CNOT gate \cite{Streltsov2015}. 

For a single-qubit state $\rho$, the geometric coherence $C_{\mathrm{g}}$
is related to the $\ell_{1}$-norm coherence $C$ as follows \cite{Streltsov2015}:
\begin{equation}
C_{\mathrm{g}}(\rho)=\frac{1}{2}[1-\sqrt{1-C(\rho)^{2}}].\label{eq:Cg}
\end{equation}
It is now crucial to note that the same functional relation holds
between the geometric entanglement $E_{\mathrm{g}}$ and the concurrence
$E$ for any two-qubit state $\mu$ \cite{Wei2003,Streltsov2010}:
\begin{equation}
E_{\mathrm{g}}(\mu)=\frac{1}{2}[1-\sqrt{1-E(\mu)^{2}}].\label{eq:Eg}
\end{equation}
Recalling that Eq.~(\ref{eq:EntCoh-2}) is fulfilled for the geometric
entanglement $E_{\mathrm{g}}$ and geometric coherence $C_{\mathrm{g}}$,
these results imply that Eq.~(\ref{eq:EntCoh-2}) also holds for
$\ell_{1}$-norm of coherence $C$ and concurrence $E$. Moreover,
for these quantifiers the CNOT gate must also be the optimal incoherent
operation, attaining equality in Eq.~(\ref{eq:EntCoh-2}). Our results
also hold if $C$ is chosen to be the trace norm coherence, as for
single-qubit states the trace norm coherence coindiced with the $\ell_{1}$-norm
coherence \cite{Shao2015}.
\end{document}